\documentclass[aip,graphicx]{revtex4-2}
\usepackage[english]{babel}
\usepackage[utf8]{inputenc}
\usepackage{csquotes}
\addto{\captionsenglish}{%
  
}
\usepackage{graphicx}
\usepackage{amsmath}
\usepackage{amssymb}
\usepackage{dcolumn}
\usepackage[T1]{fontenc}
\usepackage[colorlinks,citecolor=red,urlcolor=blue,bookmarks=false,hypertexnames=true]{hyperref}
\usepackage{multirow}
\usepackage{xcolor}
\usepackage{wrapfig}
\usepackage{adjustbox}
\usepackage{setspace}

\newcommand{\figlabel}[4][0.2]{\put(#2,#3){\parbox{#1\unitlength}{\centering #4}}}

\newcommand{\add}[1]{{#1}}

\begin{document}

\title{Kinetics of Shear Banding Flow Formation in Linear and Branched Wormlike Micelles}
\author{Peter Rassolov}
\author{Alfredo Scigliani}
\author{Hadi Mohammadigoushki}
\email[]{hadi.moham@eng.famu.fsu.edu}
\affiliation{Department of Chemical and Biomedical Engineering, FAMU-FSU College of Engineering, Tallahassee, FL 32310, USA.}

\date{\today}

\begin{abstract}
We investigate the flow evolution of a linear and a branched wormlike micellar solution with matched rheology in a Taylor-Couette (TC) cell using a combination of particle-tracking velocimetry, birefringence, and turbidity measurements. Both solutions exhibit a stress plateau within a range of shear rates. Under startup of a steady shear rate flow within the stress plateau, both linear and branched samples exhibit strong transient \add{shear thinning flow profiles}. However, while the flow of the linear solution \add{evolves to a} banded structure at longer times, the flow of the branched solution transitions to a curved velocity profile with no evidence of shear banding. Flow-induced birefringence measurements indicate transient birefringence banding with strong micellar alignment in the high shear band for the linear solution. The transient flow-induced birefringence is stronger for the branched system at an otherwise identical Wi. At longer times, the birefringence bands are replaced by a chaotic flow reminiscent of \add{elastic} instabilities. Visualization of the flow-induced turbidity in the velocity gradient-vorticity plane reveals quasi-steady banding with a turbidity contrast between high and low shear bands in the linear solution. However, the turbidity evolves uniformly within the gap of the TC cell for the branched solution, corroborating the non-banded quasi-steady velocimetry results. Finally, we show that while \add{elastic} instabilities in the linear solution emerge in the high shear band, the flow of branched solution at high Wi becomes unstable due to end effects, with growing end regions that ultimately span the entire axial length of the TC cell.

\end{abstract}

\maketitle

\section{Introduction and background}

When dissolved in aqueous solutions, surfactants can self-assemble into micelles of various shapes such as spheres, rods, vesicles, sponges, and disks~\cite{Cates2006,ISRAELACHVILI2011503}. Among these, wormlike micelles (WLMs), or long, flexible cylindrical micelles, can form entangled networks that introduce viscoelasticity to the solutions. WLM solutions are easy to prepare and unlike polymer solutions are not susceptible to shear mechanical degradation, and therefore, these systems have attracted much interest both in industrial operations as well as scientific research\cite{YANG2002276}. 

In micellar solutions that are based on a surfactant and a salt, a solution prepared with zero salt concentration typically forms spherical micelles if the surfactant concentration is beyond a critical value known as the critical micelle concentration. Addition of a salt or a counterion screens out the repulsive interactions between charged surfactant head groups, which favors the formation of cylindrical micelles. As the salt concentration is increased, the cylindrical micelles grow in length and become entangled. These entanglements form a network that causes the rheological properties of the solution (e.g., zero-shear-rate viscosity $\eta_0$) to increase. This marks the transition from the dilute to the semi-dilute concentration regime. Further increasing the salt concentration causes $\eta_0$ to continue to increase as the micelles become more entangled. However, some WLM systems exhibit a peak in $\eta_0$ at a critical salt concentration in the semi-dilute regime~\cite{Rehage1988,Candau1993,Khatory1993,In1999,Koehler2000,Raghavan2001,Croce2003}. Direct cryo-TEM images have revealed that this transition for several micellar systems is associated with a microstructural transition from linear to branched structures\cite{Croce2003,Kuperkar2008,Ziserman2009,Calabrese2015,Gaudino2015}. Over the last decade, researchers have developed a range of additional experimental methods to distinguish linear from branched WLMs such as neutron spin echo (NSE)~\cite{Cala18}, NMR\cite{Holder2021}, and extensional rheology\cite{Omid19,Chella08}. It has been suggested that in branched WLMs, the branch joints can slide along the WLMs chain, unlike the branch joints that may be formed in some polymers\cite{Lequeux1992,Appell1992}, and this sliding introduces a new mechanism for stress relaxation. This new relaxation mechanism becomes stronger at higher salt concentration (as more branched points are created), and consequently, $\eta_0$ decreases~\cite{Lequeux1992,Calabrese2015}.

Under steady shear flow conditions, some viscoelastic WLM solutions (and many other types of soft materials) may develop shear banding, or the formation of distinct regions within the flow field with different local shear rates~\cite{Divoux2015}. Shear banding in WLMs is usually associated with a shear stress plateau spanning a range of applied shear rates in the flow curve\cite{Hu2005,Callaghan2008,Manneville2008}, and may develop both in linear and branched WLM solutions~\cite{Gaudino2017}. Previous studies have suggested that the type of micellar microstructure (linear versus branched) may significantly alter the microstructural origin of shear banding~\cite{Thareja2011,Caiazza2019}. For example, in a range of linear WLMs that are in the semi-dilute concentration regime, the formation of shear banding is associated with the micellar alignment in the high shear band~\cite{Hu2005,Helgeson2009,Gurnon2014,ArenasGomez2019}. However, Thareja et al.~\cite{Thareja2011} probed shear banding flows of branched solutions of erucyl bis-(hydroxyethyl) methylammonium chloride (EHAC)/sodium salicylate(NaSal) using small-angle light scattering and found evidence of shear-induced phase separation in the high shear band. More recently, Caiazza et al.\cite{Caiazza2019} detected flow-induced concentration gradients in shear banding flows of a branched micellar solution based on cetyltrimethylammonium bromide (CTAB)/NaSal in a microfluidic channel. Calabrese et al.~\cite{Calabrese2015} studied solutions of linear and branched micellar solutions based on cetyltrimethylammonium tosylate (CTAT), sodium dodecylbenzene sulfonate (SDBS) and sodium tosylate (NaTos) using a combination of bulk rheology and rheo-SANS in a Taylor-Couette (TC; flow between two concentric cylinders) cell. These researchers found that the branched WLM solution does not exhibit a shear stress plateau, and while rheo-SANS patterns feature evidence of fluid microstructure banding for linear WLMs, no banding is reported for branched WLM solutions\cite{Calabrese2015}. Micellar branching may also modify the dynamic flow behavior of the WLM solutions in complex flows (e.g., microfluidic channels~\cite{Hwang2017,Zhang2018}, and in flow past a falling sphere~\cite{Wu2021}). \par  

While there are indications that micellar branching might affect the shear banding mechanism in TC flows or flow behavior in complex flow geometries, our understanding of the impacts of micellar branching on the evolution of shear banding flows of WLMs in simple shear flows is limited. In the small-gap limit, the TC flow provides a canonical geometry for studies of shear banding. Table~\ref{tbl:literature} shows a survey of literature on shear banding studies of WLM solutions in the TC cell and the reported fluid microstructure \add{for fluids where a shear stress plateau is reported}. According to the published literature, almost all hydrodynamic studies of the semi-dilute WLM solutions in TC flows have either unknown micellar topology (unclear whether linear or branched) or have focused on one type of micellar microstructure (linear or branched). \add{Direct comparisons} between the dynamic flow behavior of shear banding linear and branched WLM solutions in TC flows \add{are} scarce. In a recent study, Gaudino et al.~\cite{Gaudino2017} measured the quasi-steady flow profiles and micellar orientation of shear banding linear and branched WLM solutions in TC flows using a combination of velocimetry and rheo-SANS. \add{While there is some evidence for formation of branched micelles in this system beyond the viscosity peak~\cite{Gaudino2015}, these authors did not directly image the micellar microstructure at high salt concentrations to quantify the degree of branching~\cite{Gaudino2017}. It was suggested that at high salt concentrations and around the second viscosity peak, a highly dense network of branched micelles may form.} According to these researchers, shear banding is observed both for the linear and highly branched WLMs provided that the micellar contour length is long enough~\cite{Gaudino2017}. In addition, it was found that the quasi-steady micellar alignment is stronger in the shear banding branched micellar solutions compared to the shear banding linear WLMs at an otherwise identical Weissenberg number (Wi)~\cite{Gaudino2017}. Here, Wi is defined as Wi = $\lambda\dot{\gamma}$, where $\lambda$ and $\dot{\gamma}$ refer to the longest relaxation time and the imposed shear rate respectively. Moreover, the onset of shear banding (or the lower shear rate bound of the stress plateau in the flow curve) was shown to occur at a greater Wi for branched WLMs compared to the the linear WLM solutions\cite{Gaudino2017}.\par 

Equally important, recent experimental and theoretical studies have shown that the evolution of the flow of shear banding WLM solutions is strongly controlled by the rheological characteristics of the fluid, and in particular, associated dimensionless numbers such as the fluid elasticity\cite{Zhou2012,Zhou2014,Mohammadigoushki2019,Rassolov2020,Rassolov2022} and the solvent to solute viscosity ratio\cite{Zhou2008,Adams2011}. In the study of Gaudino et al.~\cite{Gaudino2017}, the above mentioned rheological parameters are not controlled in a systematic way and differ between the linear and branched systems studied. Hence, the reported differences in quasi-steady flow behavior and fluid structure may be associated with changes in the rheological characteristics of the WLM solutions~\cite{Gaudino2017}. In addition, there are still no quantitative comparisons between the linear and branched WLMs in terms of the transient evolution of shear banding flows following startup of a steady shear rate flow. In summary, our understanding of the effects of micellar branching on the evolution of the shear banding flows of WLMs in TC flows is limited.\par

The main objective of this study is to assess the impacts of micellar branching on the evolution of flow profiles and structure of shear banding WLM solutions with matched rheological properties. Our experiments are conducted on a linear and a branched WLM solution based on CPyCl/NaSal \add{under the start-up of steady applied shear rate flow} using a range of rheo-optical measurements in a Taylor-Couette cell. We analyze the evolution of \add{spatially resolved flow features} by a combination of particle tracking velocimetry (rheo-PTV), flow-induced birefringence (rheo-FIB), turbidity, and secondary flow visualizations.

\begin{table*}[h!]
\centering

\begin{tabular}{c c c c} 
\hline
System & Conc. (mM/mM) & Critical Salt Conc. (mM)* & {Micellar} Structure  \\ \hline
{CPyCl / NaSal} & 100/60~\cite{Britton1997,Mair1997} & 70**\cite{Zhang2018} & linear \\
\hline
\multirow{10}{3cm}{\centering CPyCl/NaSal/NaCl}
& 100/50/100\cite{Miller2007} & - &  linear \\
 & 200/100/100\cite{Miller2007} & - &  linear \\
 & 100/100/60.9\cite{Gaudino2017} & 60.9\cite{Gaudino2017} &  linear  \\
& 100/100/262\cite{Gaudino2017} & 60.9\cite{Gaudino2017} &  branched  \\ 
& 140/70/200\cite{Cheng2017} & - & unknown \\
& 143/71/500\cite{Salmon2003,Lettinga2009,Alkaby2018,Alkaby2020} & - & unknown \\ 
 & 174/87/50\cite{Hu2005} & $\approx$ 100\cite{Rehage1988}** &  linear  \\
 & 177/88.5/500\cite{Gurnon2014} & $\approx$ 100\cite{Rehage1988}** &  linear  \\
 & 200/120/500\cite{Britton1999} & - &  unknown \\
 & 238/119/500~\cite{LopezGonzalez2004,LopezGonzalez2006,Lettinga2009,Feindel2010,Fardin2012a} & - &  unknown  \\ 

\hline

CTAB / NaNO$_3$ & 400/405\cite{Brown2011, Lerouge2008,Mohammadigoushki2016} & - &  unknown \\

\hline
\multirow{7}{3cm}{\centering CTAB/NaSal}
& 490/0~\cite{Helgeson2009} & - & unknown \\ 
 & 50/100~\cite{Decruppe2006} & 25-50\cite{Rothstein03} & branched \\
 & 9/9\cite{Mohammadigoushki2019} & - &  linear \\
 & 10/9, 12/10.5, 15/12~\cite{Rassolov2020} & 9.5, 11, 12.5\cite{Rassolov2020} & linear  \\ 
 & 20/15, 25/17.5, 50/27~\cite{Rassolov2022}& 16, 18.5, 30~\cite{Rassolov2022} & linear \\
 & 100/40, 200/56~\cite{Rassolov2022}& 45, 60~\cite{Rassolov2022} & linear  \\
 & 150/67.5, 200/50, 200/70~\cite{Rassolov2022} & 72, 75~\cite{Rassolov2022} & linear  \\
 \hline

\multirow{2}{3cm}{\centering CTAT/SDBS/NaTos}
& 1.455/0.045/0.01~wt\% \cite{Calabrese2015} & - &  linear  \\

 & 1.455/0.045/0.05~wt\% \cite{Calabrese2016} & - &  branched  \\ \hline
EHAC / NaSal & 40/300, 40/400, 40/600~\cite{Thareja2011} & 250 \cite{Thareja2011} &  branched  \\ 
\hline 
TDPS/SDS/NaCl & 46/25.3/200~\cite{ArenasGomez2019}& 200~\cite{ArenasGomez2019} &  linear \\
\hline
\end{tabular}
\caption{Summary of previous studies of shear banding linear and branched WLM solutions in the semi-dilute concentration regime in the TC flow. *Salt concentration where the linear-to-branched transition is reported, or where $\eta_0$ is at the reported local maximum. **Approximated by extrapolating the reported range of salt and surfactant concentrations. }
\label{tbl:literature}
\end{table*}

\section{Experiments}

\subsection{Materials}

The wormlike micellar solutions of this study were made by mixing Cetylpyridinium Chloride (CPyCl) and NaSal in de-ionized water, stirred until uniformly mixed, and left to equilibrate for a minimum of 2 weeks prior to experiments. Both CPyCl and NaSal were obtained from Millipore Sigma and used as received. To resolve the local velocity profiles, wormlike micellar fluids were prepared with 50 ppm by mass of glass microspheres (Potters 110P8); our previous studies confirmed that these particles do not change the rheology~\cite{Mohammadigoushki2019,Rassolov2020}. In addition, to visualize the secondary flows, fluids were prepared with 250 ppm by mass of mica flakes (Jacquard PearlEx 671). \add{We selected the WLMs based on CPyCl/NaSal because TEM images have confirmed the transition from linear to branched structures beyond the viscosity maximum in this system~\cite{Gaudino2015,abezgauz2010}. }

\subsection{\add{Characterization methods}}

\subsubsection{\add{Bulk rheology}}
Fluids were characterized using a commercial rheometer (Anton Paar MCR 302) with off-the-shelf Taylor-Couette (TC) measuring geometry ($R_i$ = 13.328 mm, $R_o$ = 14.449 mm, $h$ = 40 mm where $R_i$ and $R_o$ are respectively the inner and outer cylinder radii and $h$ is the height of the inner cylinder). As in previous studies~\cite{Rassolov2020,Rassolov2022}, two types of measurements were used: Small Amplitude Oscillatory Shear (SAOS) was used to obtain the linear viscoelastic responses, and steady applied shear was used to obtain the flow curves. \add{Note that at this stage, where we varied surfactant and salt concentration as well as the temperature to obtain the two WLMs with different structures and matched rheology, we primarily used the off-the-shelf TC cell to characterize the rheology of WLMs. Once two WLMs were identified, we performed the remaining experiments with the rheo-optical apparatus, and the resulting rheological properties were found to be consistent with the results obtained by the off-the-shelf TC cell geometry (see Fig.~S1 in the supplementary materials).}

\subsubsection{\add{$^1$H NMR diffusometry}}

\add{Although bulk rheology measurements and in particular, zero-shear viscosity can be used as an indicator for the microstructural transition from linear to branched in CPyCl/NaSal system, direct evidence for formation of branched micelles is obtained by TEM imaging. However, as noted in our earlier study\cite{Holder2021} and pointed out by others\cite{clausen1992}, there are some caveats associated with TEM imaging of WLMs. In addition to being extremely challenging, experiments with TEM are time-consuming and especially the sample preparation stage may inadvertently change the equilibrium structure of the WLMs\cite{clausen1992}. Recently, we proposed a non-invasive technique based on nuclear magnetic resounance (NMR) diffusometry that is highly sensitive to the type of micellar microstrucuture (linear vs. branched)\cite{Holder2021}. Therefore, to examine the type of micellar microstructures below and above the viscosity peak (and the extent of branching beyond the viscosity peak), we performed $^{1}$H NMR diffusometry via a 21.1-T wide bore magnet equipped with a 900-MHz ultrawide bore spectrometer that uses a Bruker AVIII console capable of performing liquid-state, solid-state and microimaging analysis. This magnetic system has Bruker Micro2.5 (40-mm ID) microimaging gradients and a widebore (64-mm ID) RRI gradient set coupled to 60-A Great60 amplifiers. WLMs were placed in sealed 5 mm NMR capillary tubes and measurements were performed in $^{1}$H linear birdcage. Similar to our previous study\cite{Holder2021}, the diffusion weighted measurements were performed using a pulsed gradient stimulated echo sequence (see more details about the sequence in~\cite{Holder2021}).}\par 
\add{In principle, the signal attenuation in diffusion weighted NMR spectroscopy can be defined as: 
\begin{equation}
    \frac{S(q)}{S(0)} = \exp \left(-\frac{1}{2}{Z}^{2}q^{2}\right),  
\end{equation}
where $S$ denotes the strength of the NMR signal intensity, $q$ is the diffusion weighting, and $S(0)$ is the NMR signal with no diffusion ($q = 0$). The diffusion weighting $q$ is defined as $q = \gamma \zeta g$, where $\gamma$ is the gyromagnetic ratio of the proton ($\gamma = 2.67\times 10^8$ rad/s/T), $g$ is the strength of the magnetic gradient pulse, and $\zeta$ is the gradient pulse duration time. In addition, $Z$ is the mean square displacement of the molecule and can be expressed as: $Z\propto (T_{diff})^{\alpha}$ with $\alpha$ being a power-law index that can be linked to the types of self-diffusion. For example, for free and unrestricted self-diffusion of molecules in the bulk (or Brownian diffusion), $\alpha=$1\cite{tanner1968}. For restricted diffusion, $\alpha$ decreases approaching $\alpha= $1/2, which characterizes curvilinear diffusion of molecules in tubes. Holder and co-workers have previously shown that depending on the types of micellar structure, $\alpha$ can vary between 0.5 (for linear) and 1 (for highly branched) wormlike micellar solutions~\cite{Holder2021}. Therefore, we will use this indicator ($\alpha$ exponent and diffusion type) to assess the extent of branching in our WLMs. In the experiments reported in this paper, $T_{diff}=$ 15-500 milliseconds and $g=$ 0 - 372.1 mT/m with a fixed gradient pulse duration ($\zeta$) of 6 ms. The acquired NMR data were processed using Bruker TopSpin 4.1.4 software. A 10 Hz filter was used to apodise the time-domain data and spectra were phase-corrected, and analyzed using both peak intensity as a function of diffusion weighting for each diffusion time.}

\subsection{Rheo-optical methods}

Once fluids were characterized and a matching pair was identified, the selected fluids were visualized under startup of a steady shear flow using a custom-built rheo-optical apparatus described in our previous work~\cite{Rassolov2020,Rassolov2022} with $R_i$ = 13.35 mm, $R_o$ = 14.53 mm, and $h$ = 50 mm. \add{In this apparatus, temperature was controlled by circulating water through a bath surrounding the flow cell using a temperature control circulator/chiller. The temperature was measured at several points in the water bath surrounding the rheo-optical cell and was found to be uniform throughout the geometry and steady over time within the accuracy of the thermometer (about $\pm 0.2 ^{\circ}\mathrm{C}$).} As with the more recent work~\cite{Rassolov2022}, the inner cylinder was roughened to reduce wall slip at its surface. This apparatus was used for each of four spatially resolved fluid measurements as described below and shown in Fig.~\ref{fig:rheoptv-apparatus}. \add{During each measurement, the evolution of shear stress was recorded simultaneously with rheo-optical data acquisition. Hence, in all reported shear stress vs. strain data in the manuscript, we have used the rheo-optical cell.}

\subsubsection{Rheo-PTV}

\begin{figure*}[t]
    \centering
    \setlength{\unitlength}{14cm}
    \begin{picture}(1,0.5)
    \put(0,0){\includegraphics[width=\unitlength]{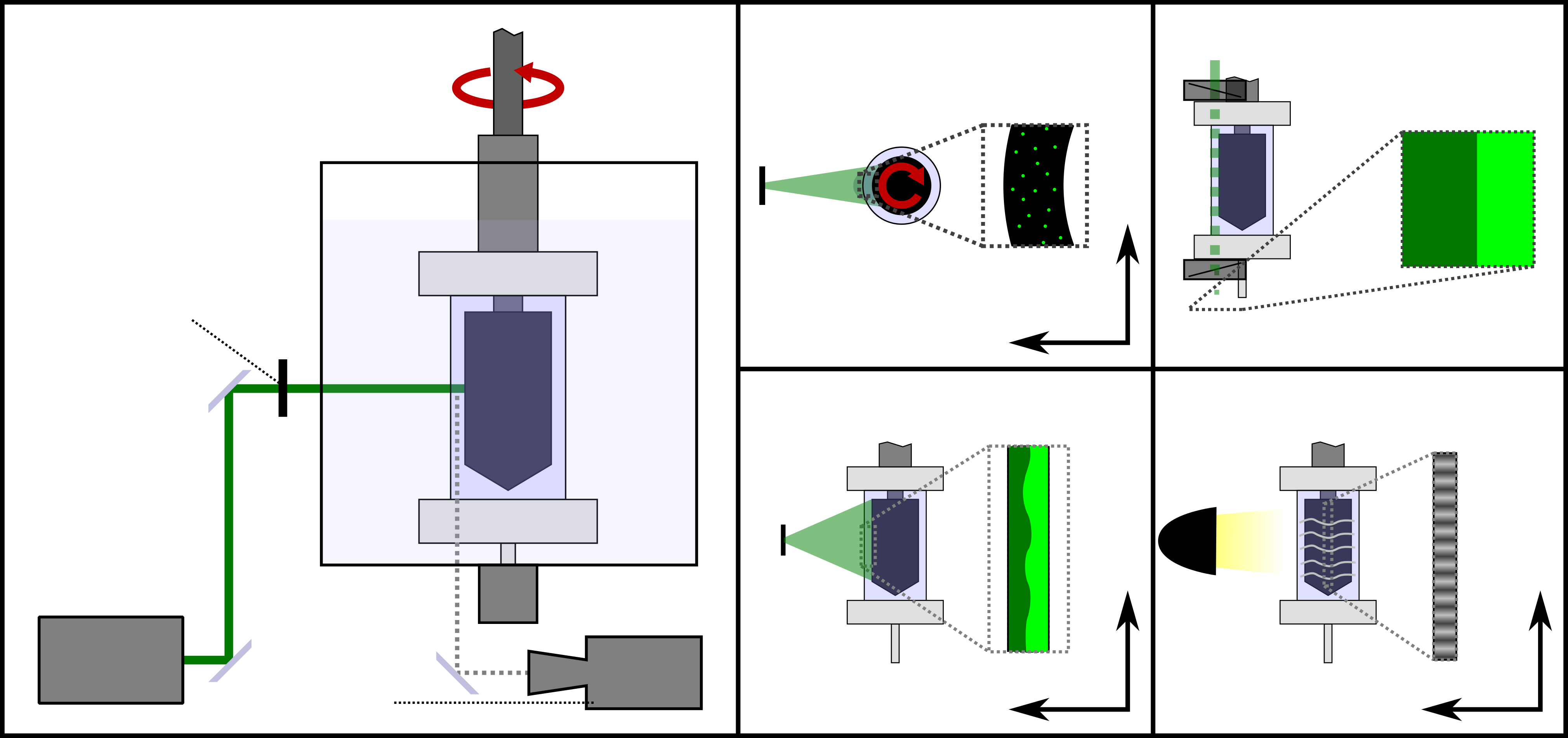}}
    \figlabel[0.1]{0.02}{0.09}{Laser}
    \figlabel[0.10]{0.02}{0.27}{Powell lens}
    \figlabel{0.11}{0.02}{Camera}
    \figlabel[0.05]{0.60}{0.25}{$x_2$}
    \figlabel[0.05]{0.69}{0.34}{$x_1$}
    \figlabel[0.05]{0.60}{0.01}{$x_2$}
    \figlabel[0.05]{0.69}{0.10}{$x_3$}
    \figlabel[0.05]{0.87}{0.01}{$x_1$}
    \figlabel[0.05]{0.96}{0.10}{$x_3$}
    \figlabel[0.05]{0.00}{0.44}{(a)}
    \figlabel[0.25]{0.48}{0.44}{(b) Rheo-PTV}
    \figlabel[0.25]{0.74}{0.44}{(c) Rheo-FIB}
    \figlabel[0.25]{0.48}{0.21}{(d) Turbidity}
    \figlabel[0.25]{0.74}{0.21}{(e) Secondary flows}
    \end{picture}
    \caption{(a) A schematic of the rheo-optical cell apparatus. (b-e) Schematics of illumination and imaging of rheo-optical experiments: (b) rheo-PTV, (c) rheo-FIB, (d) turbidity visualizations, and (e) secondary flows visualizations. \add{Directions in the images are indicated: $x_1$, the direction of flow; $x_2$, along the shear stress gradient (spanning the gap); $x_3$, the vorticity direction (parallel with the concentric cylinder axis).} (Not to scale.)}
    \label{fig:rheoptv-apparatus}
\end{figure*}

Spatially and temporally resolved fluid velocity data were obtained using steady shear rheometry with particle tracking velocimetry (rheo-PTV; Fig.~\ref{fig:rheoptv-apparatus}(b)) as described in our previous work~\cite{Rassolov2020}. A laser beam was diverged to illuminate the fluid in a plane orthogonal to the TC cell rotation axis (i.e. the $x_1$-$x_2$ plane), and the fluids were seeded with glass microspheres as tracking particles. \add{We confirmed that adding these particles did not change the rheology of our solutions (see Fig.~S1 in the supplementary materials).} A high-speed video camera (Phantom Miro 310) focused on the illuminated plane was used to record video of the particles moving with the fluid as the flow was applied. The PTV analysis was completed on fluid near the center of the TC cell along its axis at $X_3  = 25 \pm 5$ mm, where $X_3 = 0$ and $X_3 = 50$ mm refer to the bottom and the top of the inner cylinder, respectively. The video was processed using the Python script in our previous work~\cite{Rassolov2022} based on TrackPy~\cite{Allan2019} to obtain time-dependent local velocity data in an array of space-time elements. In addition, linear least-squares regression was used to estimate the time-dependent local shear rate from the velocity data. The local shear rate data were used to quantify shear rate inhomogeneity in the TC cell using the shear rate drop parameter $\Delta$~\cite{Adams2011}:
\begin{equation}
    \Delta = \frac{\dot{\gamma}_{max}-\dot{\gamma}_{min}}{\dot{\gamma}_{app}},
    \label{eqn:Delta}
\end{equation}
where $\dot{\gamma}_{max}$, $\dot{\gamma}_{min}$, and $\dot{\gamma}_{app}$ are the maximum, minimum, and applied shear rates in the TC cell, respectively. \add{The shear rate for a unidirectional, rectilinear flow field is given by $\dot{\gamma} = \partial v / \partial x_2$ and is applicable as an approximation to TC flow if the curvature is small. However, quantifying the local shear rate for a strongly inhomgoeneous flow field requires computing a numerical derivative of measured data over narrow domains, which greatly amplifies measurement noise. Therefore, two methods were used to obtain $\dot{\gamma}$ experimentally. First, for non-banded velocity profiles, linear least-squares fits to the data in small intervals of the flow field were used. Second, where applicable, a sigmoid-based function representing a two-banded velocity profile was fitted to the velocity data throughout the entire flow field (the specific function, fitting procedure, and sample fitted data \add{Fig.~S2} are given in the supplementary materials). In addition to these two methods, smoothing by low-order localized polynomial fitting was used to obtain numerical derivatives for the purpose of distinguishing shear banding from shear thinning. This method is described in detail in prior work by Cheng et al.\cite{Cheng2017}, where it was used to compute derivatives of velocity profiles up to the third order. In this work, we apply this method with a third order polynomial, smoothing parameter $p = 0.2$ (i.e. 20\% of the data set was used as the subset for each fit), and no weighting.}

\subsubsection{Flow-induced birefringence}

Alignment of the wormlike micelles under shear flow causes birefringence~\cite{Kim2000,Helgeson2009}. Flow-induced birefringence (FIB) was measured using a pair of crossed polarizing filters as shown schematically in Fig.~\ref{fig:rheoptv-apparatus}(c). For FIB measurements, a light beam was passed through the polarizer, then it propagated through the fluid along the $x_3$ axis, and then through the analyzer, which is always orthogonal to the polarizer. If the light used is monochromatic and flow-induced dichroism is negligible, the birefringence can be determined quantitatively using the method proposed by Osaki et al.~\cite{Osaki1979,Wu2021}. In this method, the experiment trial was repeated for two different orientations of the polarizer: once with the polarizer parallel to the flow ($I_0$), and once with the polarizer at 45 degrees ($I_{45}$). The transmitted light intensity for each of these orientations is related to the birefringence as follows:
\begin{subequations}
\begin{align}
    I_0 = I^\prime \sin^2 (2\chi) \sin^2 \left( \frac{\delta}{2} \right) \label{eqn:i0} \\
    I_{45} = I^\prime \cos^2 (2\chi) \sin^2 \left( \frac{\delta}{2} \right), 
    \label{eqn:i45}
\end{align}
\end{subequations}
where $I^\prime$ is the intensity of the light in the absence of the analyzer, $\chi$ is the extinction angle (or the orientation of the refractive index tensor relative to the $x_1$ direction), and $\delta$ is the phase difference in the light wave due to birefringence. These equations can be combined and rearranged to obtain $\chi$ and $\delta$:
\begin{subequations}
\begin{align}
    \chi = \frac{1}{2} \tan^{-1} \left( \sqrt{\frac{I_0}{I_{45}}} \right) \label{eqn:chi_arctan} \\
    \delta = \sin^{-1} \left( \sqrt{\frac{I_0 + I_{45}}{I^{\prime}}} \right). \label{eqn:delta_arcsin}
\end{align}
\end{subequations}
Although $\chi$ can be defined as the smallest angle between a refractive index tensor principal axis and the $x_1$ direction meaning that Equation~\ref{eqn:chi_arctan} is always valid, Equation~\ref{eqn:delta_arcsin} requires the assumption $\delta < \pi/2$. Finally, $\delta$ can be related to the birefringence intensity $\Delta n^\prime$ as:
\begin{equation}
    \Delta n^\prime = \frac{\delta \Lambda}{2 \pi H},
\end{equation}
where $\Lambda$ is the wavelength of the incident light and $H$ is the optical depth of the birefringent sample.\par 

For FIB measurements, the rheo-optical cell has been modified in two ways. First, the glass outer cylinder was replaced by an opaque plastic (Delrin) cylinder in order to eliminate light passing through it and to reduce light reflection from the surface. Second, in order to eliminate residual birefringence in the end caps due to internal stresses from manufacturing and assembly, holes were drilled in the caps at the location of the light path, and 220~$\mu$m thick glass windows were installed. The resulting assembly was checked for residual birefringence in the light path by holding it between crossed polarizing films with no fluid inside. The light beam used is the same laser used for rheo-PTV experiments, which has a wavelength of 532 nm. The beam is expanded by a factor of 5 to obtain as nearly uniform $I^{\prime}$ as possible throughout the gap of the TC cell. After passing through the analyzer, the light beam falls on a screen, and the high-speed video camera is used to capture the image formed by the light on the screen.

\subsubsection{Turbidity visualization}

Under shear banded flows, some WLM solutions exhibit turbidity differences between the high and low shear bands in $x_2-x_3$ (vorticity gradient) plane of the TC cell~\cite{Mohammadigoushki2016,Fardin2012a}. To assess the significance of such turbidity contrasts in linear and branched wormlike micellar solutions, the $x_2-x_3$ plane of the TC cell was illuminated with the diverged laser beam (Fig.~\ref{fig:rheoptv-apparatus}(d)). In addition, to minimize the effects of scattered light from the inner cylinder, the video was recorded with a strong gamma adjustment. The turbidity visualizations were completed in a small window with a height of 5.1~mm that is centered 25 mm from the bottom of the TC cell.

\subsubsection{Flow visualization with mica flakes}

To identify the emergence of secondary flows and \add{elastic} instabilities, the rheo-optical apparatus was loaded with fluid prepared with mica flakes, which orient towards flow direction and reflect light in an orientation-dependent manner. \add{As with the rheo-PTV tracking particles, we confirmed that adding the mica flakes did not change the rheology of our solutions (see Fig.~S1 in the supplementary materials).} The fluid was illuminated by a desk lamp, and the flow was recorded in the $x_1$-$x_3$ plane using the same video camera. Recorded videos were assembled into an image showing spatiotemporal evolution of the flow using a Python script as in our previous work~\cite{Rassolov2022}.


\section{Results and discussions}
\subsection{Bulk rheology}
\sloppy

\begin{figure}[htb]
    \centering
    \includegraphics{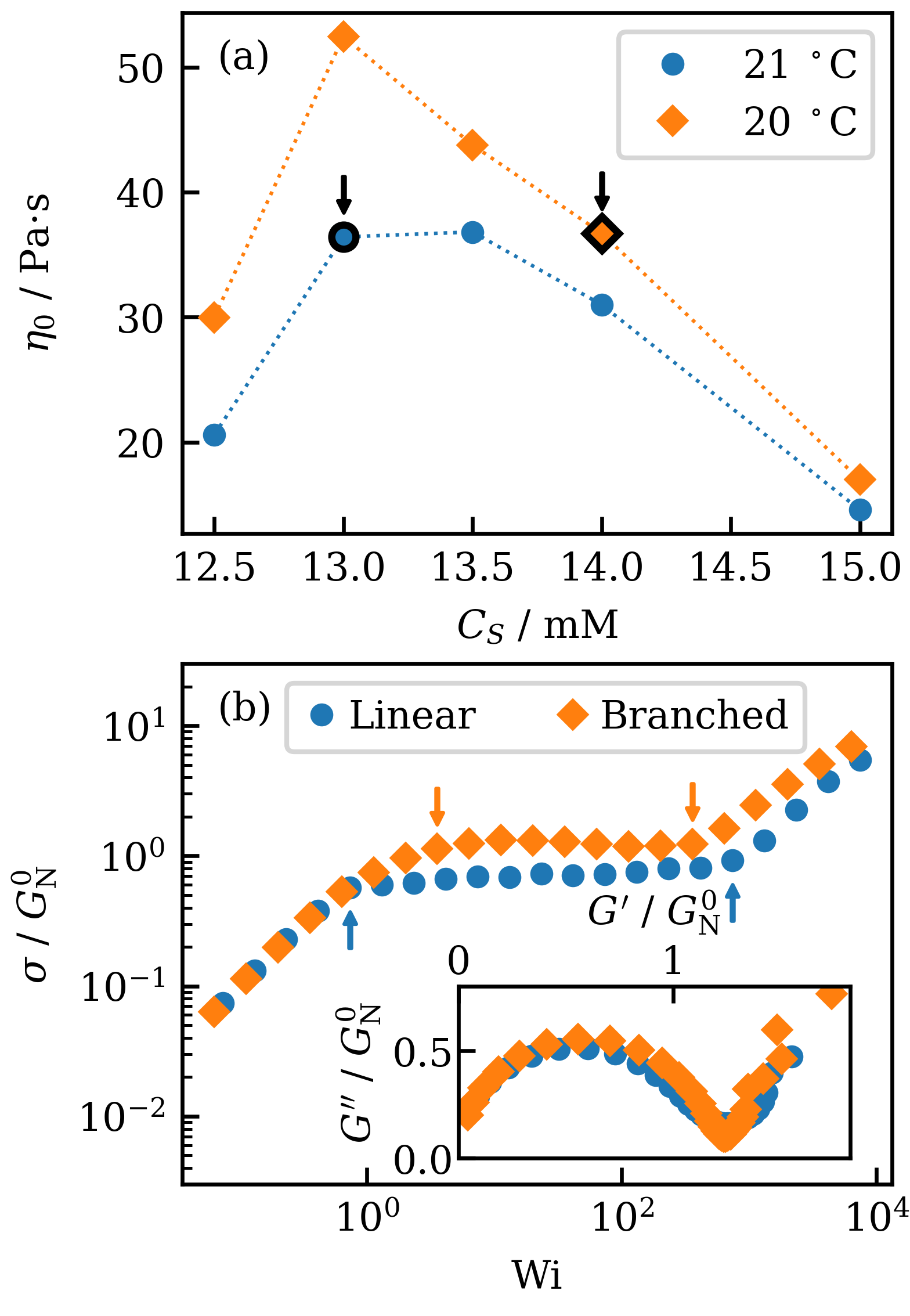}
    \caption{(a) Zero-shear-rate viscosity of the micellar solutions of CPyCl/NaSal as a function of salt concentration (C$_S$) at two different temperatures. (b) Quasi-steady dimensionless stress as a function of imposed Wi for the linear and branched micelles selected for this study. G$_N^0$ denotes the plateau modulus obtained from SAOS experiments. High and low limits of the stress plateau are indicated with arrows and are as follows: linear WLMs, Wi = 0.7 to 740; branched WLMs, Wi = 3.6 to 360. The inset of part b shows the Cole-Cole representations of the linear viscoelasticity of the two wormlike micellar solutions.}
    \label{fig:rheoresults}
\end{figure}

Fig.~\ref{fig:rheoresults}(a) shows the zero shear rate viscosity of a series of surfactant solutions based on CPyCl/NaSal at different salt to surfactant concentration ratios and temperatures. As expected, this rheological property shows a non-monotonic behavior as a function of salt to surfactant concentration ratio, with a peak that marks the transition from linear to branched WLMs. To achieve a linear and a branched micellar solution with matched rheological properties, we selected a linear WLM solution of CPyCl/NaSal (15~mM/13~mM) at 21 $^\circ$C, and a branched WLM solution of 15~mM/14~mM at 20 $^\circ$C. Measured flow curves indicate that these micellar solutions exhibit a stress plateau within a range of shear rates (see Fig.~\ref{fig:rheoresults} (b)). Although the rheological properties (e.g., $\eta_0$ and $\lambda$; see Table~S1 and Fig.~S3 of the supplementary materials) of these two systems are similar, the onset and the end of stress plateau vary between these two samples. In particular, the onset of stress plateau occurs at a larger Wi in the branched system compared to the linear micellar solution, in agreement with the previous observations of Gaudino et al.~\cite{Gaudino2017}. Additionally, the inset of Fig.~\ref{fig:rheoresults}(b) shows a semi-circular Cole-Cole plots for these two micellar solutions indicating that these systems are in the fast-breaking regime. \add{Although based on bulk rheology data, the data of Fig.~\ref{fig:rheoresults}(a) suggests that WLM solution of CPyCl/NaSal (15~mM/14~mM) at 20 $^\circ$C forms branched structure, the extent of branching (or branching density) cannot be accessed by the bulk rheology data. Therefore, to assess the extent of branching in this system, we have performed $^{1}$H NMR diffusometry experiments detailed below.}\par 

\subsection{\add{$^1$H NMR diffusometry}}

\begin{figure}[htb]
    \centering
    \includegraphics{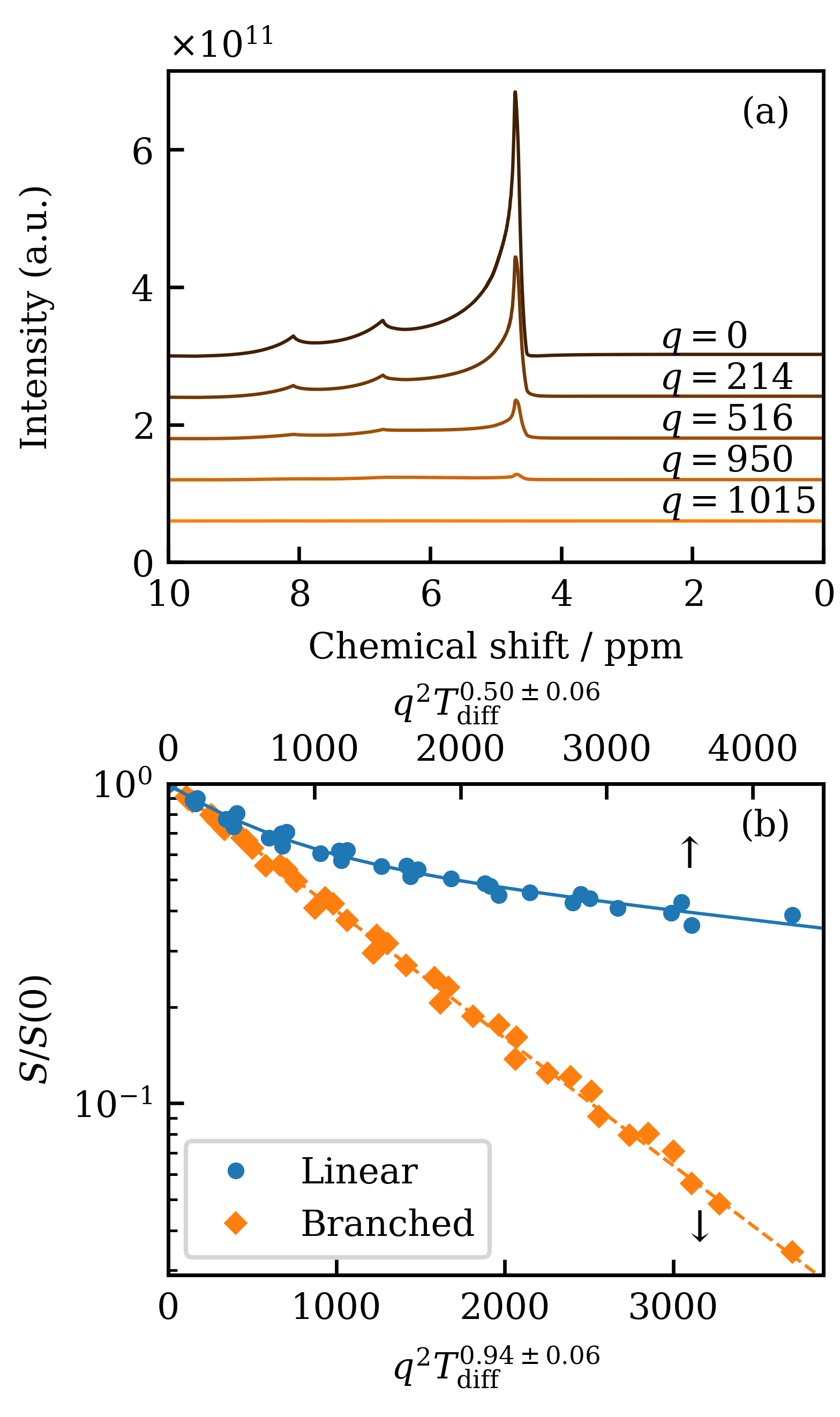}
    \caption{\add{(a) Raw NMR signal intensities as a function of chemical shift for CPyCl/NaSal (15~mM/14~mM) at 20 $^\circ$C for various diffusion weighting $q$ values. The unit of diffusion weighting q is [1/mm]. (b) Normalized NMR signal intensity of the resonance associated with the surfactant molecule (at a chemical shift 6.6 ppm) as a function of diffusion weighting. The diffusion scaling with diffusion time ($T_\mathrm{diff}$) is shown for branched WLMs (bottom x-axis) and linear WLMs (top x-axis).}}
    \label{fig:NMR}
\end{figure}

\add{ Fig.~\ref{fig:NMR}(a) shows the NMR spectra as a function of frequency for the CPyCl/NaSal (15~mM/14~mM) at 20 $^\circ$C for various diffusion weighting $q$ values. Consistent with our previously published work on CPyCl/NaSal solution, the peak at a chemical shift of 4.7 is associated with the solvent (de-ionized water in this case) and the remaining resonances at larger chemical shifts are associated with salicylate ion that is incorporated in the micellar structure\cite{Holder2021}. To ensure the accuracy of our measurements, we have analyzed the signal attenuation of water molecules in linear and branched WLMs and obtained the diffusion coefficient of the water molecules (see Fig. S4 in the supplementary materials). The resulting diffusion coefficient is consistent with the reported values in the literature\cite{Holz00,tofts2000}. Subsequently, we performed a similar analysis of other chemical shifts associate with the surfactant molecule.\par 

Fig.~\ref{fig:NMR}(b) shows the measured NMR signal attenuation as a function of diffusion weighting for the two WLMs. Clearly, there is a stark difference between the surfactant self-diffusion mechanism in these two wormlike micellar solutions. While at lower salt concentration, signal attenuation exhibits a restricted diffusion behavior (with $\alpha=$ 0.5$\pm$0.06) that deviates significantly from a mono-exponential type decay, the micellar solution at higher salt content, clearly follows a mono-exponential signal decay with $\alpha=$ 0.94$\pm$0.06. These results confirm our hypothesis that for the solution with low salt concentration, the wormlike micelles are linear and at higher salt concentration, micelles form branched structures. Holder et al.~\cite{Holder2021} showed that while for moderately branched wormlike micelles, the NMR signal decay deviates from the mono-exponential type, for highly branched network of micelles (at high salt to surfactant concentration ratios), the NMR signal decay exhibits a mono-exponential behavior with $\alpha\approx$1. Therefore, we conclude that the extent of branching in the WLMs with a higher salt concentration is quite substantial.} 

\subsection{Visualization of flow profiles via Rheo-PTV}

\begin{figure}[htp]
    \centering
    \includegraphics{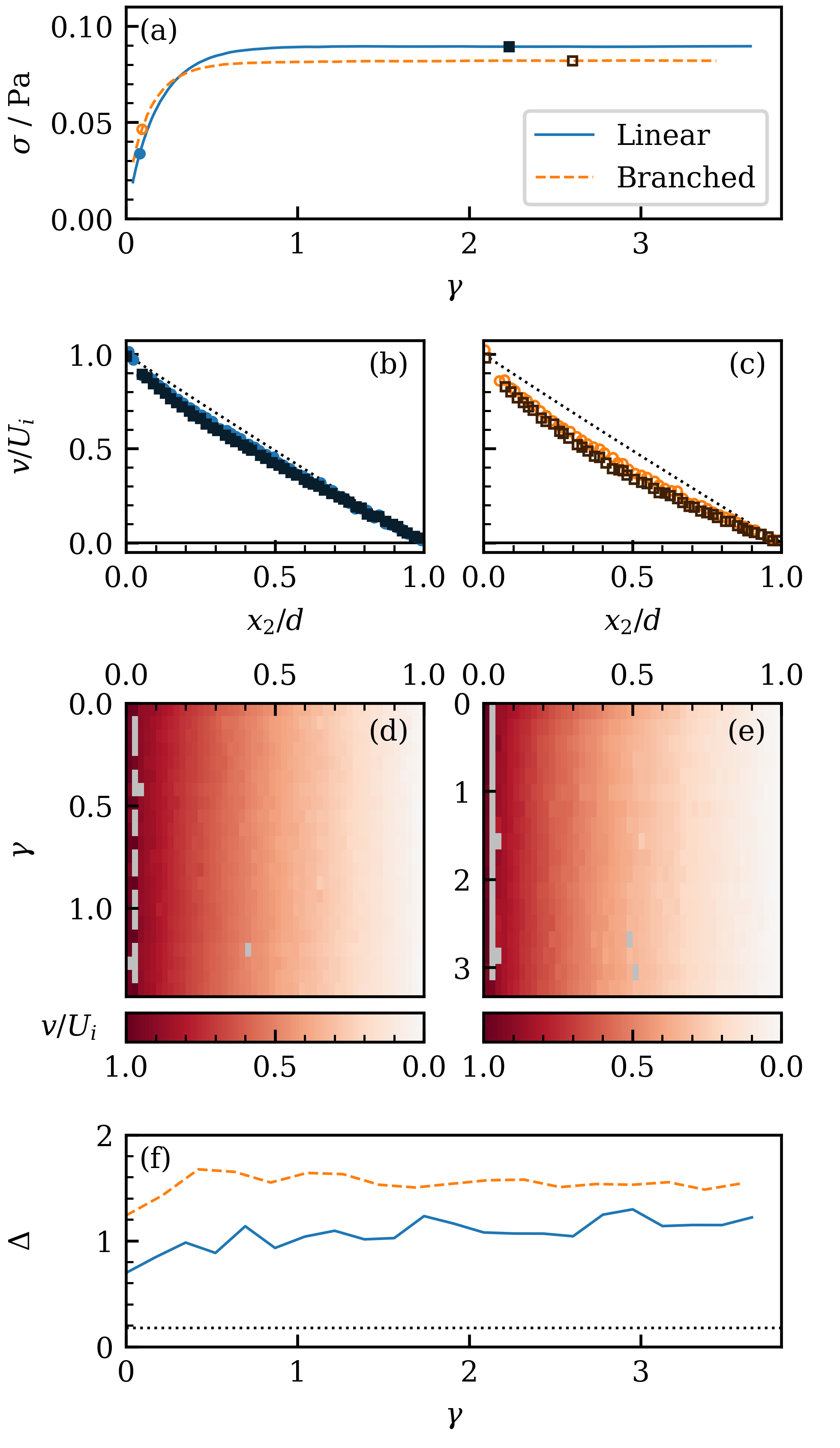}
    \caption{Local velocimetry of linear and branched WLMs at an applied Wi = 0.2: (a)~shear stress evolution with shear strain for linear and branched micellar solutions; (b,c)~selected velocity profiles for (b)~linear and (c)~branched WLMs \add{with shear strain indicated by the matching symbol in (a)}; (d,e)~maps of local velocity evolution with shear strain for linear (d) and branched (e) micellar solutions; (f)~evolution of the shear rate inhomogeneity parameter $\Delta$ with shear strain. Dotted black lines in (b,c,f) are the theoretical velocity profiles and $\Delta$ for a Newtonian fluid in the TC cell used in this study.}
    \label{fig:ptv0}
\end{figure}

To assess the impact of micellar microstructure on the local flow fields (i.e., spatio-temporal evolution of velocity profiles), we have performed startup of steady shear flow at different applied Wi values throughout the flow curve; specifically, flow visualization experiments were completed below, within, and beyond the shear stress plateau region. Fig.~\ref{fig:ptv0} shows the evolution of velocity profiles for the linear and branched wormlike micelles at Wi = 0.2. At this Wi, which is below the onset of the stress plateau in both linear and branched systems, the shear stress does not show any overshoot and gradually approaches the steady state value for both systems (see Fig.~\ref{fig:ptv0}(a)). Additionally, the measured velocity profiles of Fig.~\ref{fig:ptv0}(b,c) show that wall slip is negligible and the evolution of velocity profiles are similar for these two systems. Interestingly, the branched micelles show a steady state curved velocity profile that deviates more from the \add{theoretical} Newtonian flow compared to the linear micellar solution. To further quantify the degree of inhomogeneity in velocity profiles, we estimated the shear rate inhomogeneity parameter $\Delta$ for these two solutions at Wi = 0.2. 
Fig.~\ref{fig:ptv0}(f) shows that $\Delta$ for branched micelles is slightly higher than the linear micellar solutions. Included in this plot is also the analytical value of $\Delta$ for a Newtonian fluid (dashed line in Fig.~\ref{fig:ptv0}(f)). Interestingly, the degree of inhomogeneity in the branched micellar solution is higher than in the linear solution. Note that at this Wi, both systems are in \add{a Newtonian-like flow regime} as indicated in Fig.~S3 of the supplementary materials. The deviations from \add{a Newtonian-like} profile are presumably due to the micellar alignment in the flow. The latter results suggest that the branched micelles may align more strongly in flows compared to the linear micelles even below the onset of the stress plateau. In fact, this is consistent with the neutron scattering based findings of Gaudino et al.~\cite{Gaudino2017} that showed branched micelles are more likely to align in the flow direction at Wi values below the onset of the shear stress plateau.
\begin{figure}[htp]
    \centering
    \includegraphics{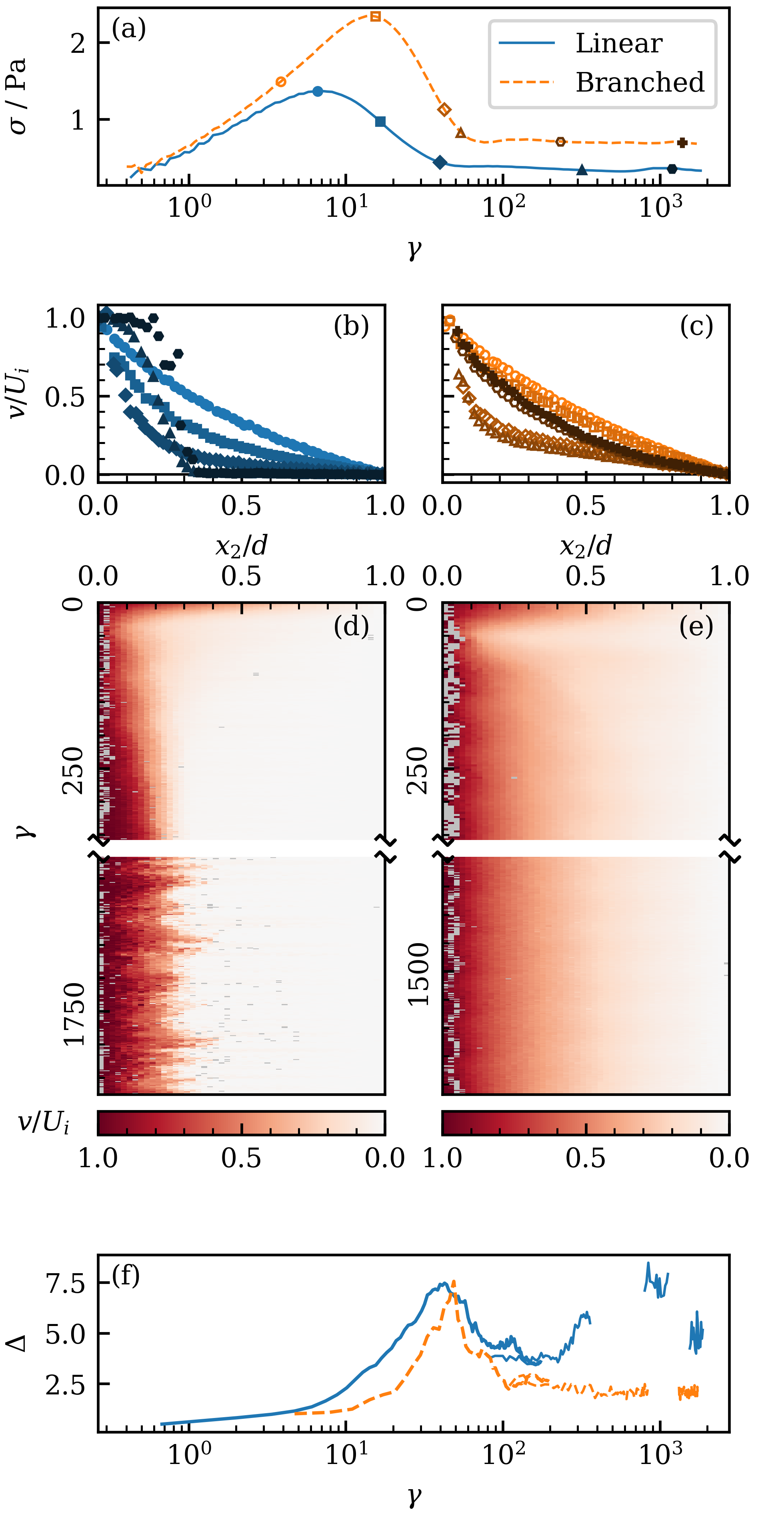}
    \caption{Local velocimetry of linear and branched WLMs at applied Wi = 50: (a)~shear stress evolution with shear strain; (b,c)~selected velocity profiles for (b)~linear and (c)~branched WLMs \add{with shear strain indicated by the matching symbol in (a)}; (d,e)~maps of local velocity evolution with shear strain \add{for linear (d) and branched (e) micellar solutions}; (f)~evolution of the shear rate inhomogeneity parameter $\Delta$ with shear strain. In subfigure (f), broader lines show $\Delta$ computed by sigmoidal shear banding fitting, while narrower lines show $\Delta$ by piecewise linear fitting.}
    \label{fig:ptv50}
\end{figure}
Subsequent to experiments at low Wi, we performed rheo-PTV experiments for Wi within the stress plateau region of the flow curves. Fig.~\ref{fig:ptv50} shows the evolution of velocity profiles for the linear and branched micellar solutions at Wi = 50. As expected, the evolution of the bulk shear stress as a function of shear strain for these two solutions is similar (see Fig.~\ref{fig:ptv50}(a)). Fig.~\ref{fig:ptv50}(b,c) show the temporal evolution of the velocity profiles for the linear and branched systems at different points along the stress curve of Fig.~\ref{fig:ptv50}(a). As reported in our previous work for linear WLMs, the flow is initially nearly uniform as the shear stress increases. The shear stress goes through a maximum around $\gamma \approx 7$ and begins to decay towards a steady state value, and concurrently, the velocity profile develops a curve \add{consistent with the emergence of shear thinning, with the shear rate decreasing from the inner cylinder towards the outer cylinder.} For branched WLMs, this initial evolution of \add{shear stress and velocity profiles} is similar except that the maximum in shear stress is around $\gamma \approx 15$. However, after the shear stress reaches the steady state value, marked differences between linear and branched WLMs appear in the velocity profiles. In the linear WLMs, a sharp shear banding interface develops around $x/d = 0.2$, consistent with previously reported results~\cite{Mohammadigoushki2016,Fardin2012a}. Then, a second low shear rate band emerges at the inner cylinder and grows outward. In the branched WLMs, there is initially evidence of \add{strong shear thinning} beginning to emerge near the inner cylinder when the shear stress approaches the steady state value. However, the velocity profile then evolves towards a quasi-steady broad curve with \add{much weaker shear thinning}. \add{Following Cheng and Helgeson \cite{Cheng2017}, we further analyzed the time-resolved velocity profiles by fitting the experimental data to low-order polynomials, and results are shown in Fig.~S5-6 of the supplementary materials. Clearly, features characteristic of shear banding emerge in linear WLMs as the shear stress approaches the quasi-steady state, while no such shear banding features are evident in the branched WLMs.} Additionally, Fig~\ref{fig:ptv50}(b,c) show minimal or no wall slip at both cylinders throughout the evolution of flow for both linear and branched WLM solutions (see Fig.~S7 and Fig.~S8 of the supplementary information, where the estimated wall slip is less than the measurement error.). Spatio-temporal evolution of the velocity profiles in the linear and branched WLMs are shown as a color map in Fig.~\ref{fig:ptv50}(d) and Fig.~\ref{fig:ptv50}(e), where these same trends can be observed.

Fig.~\ref{fig:ptv50}(f) shows that for the linear and branched WLM solutions, a near-uniform flow field is observed shortly after imposition of the shear (indicated by small values of $\Delta$), and then non-uniform profiles quickly emerge and intensify until $\gamma \approx 40-50$. As time progresses, the non-uniformity (or $\Delta$) decreases somewhat both for linear and branched micellar solutions. For linear WLMs, $\Delta$ stabilizes around $\gamma \approx 85$ and remains steady until it suddenly increases again around $\gamma \approx 200$. For branched WLMs, the decay in $\Delta$ is slower and remains around the steady value afterwards. Clearly the degree of inhomogeneity of shear rate in the linear WLMs is larger than the branched WLMs during the quasi-steady flow. The latter results confirm the above conclusion that while linear WLMs exhibit quasi-steady shear banded velocity profiles, branched WLMs do not show any signs of shear banding \add{in the} quasi-steady state.

\begin{figure}[htp]
    \centering
    \includegraphics{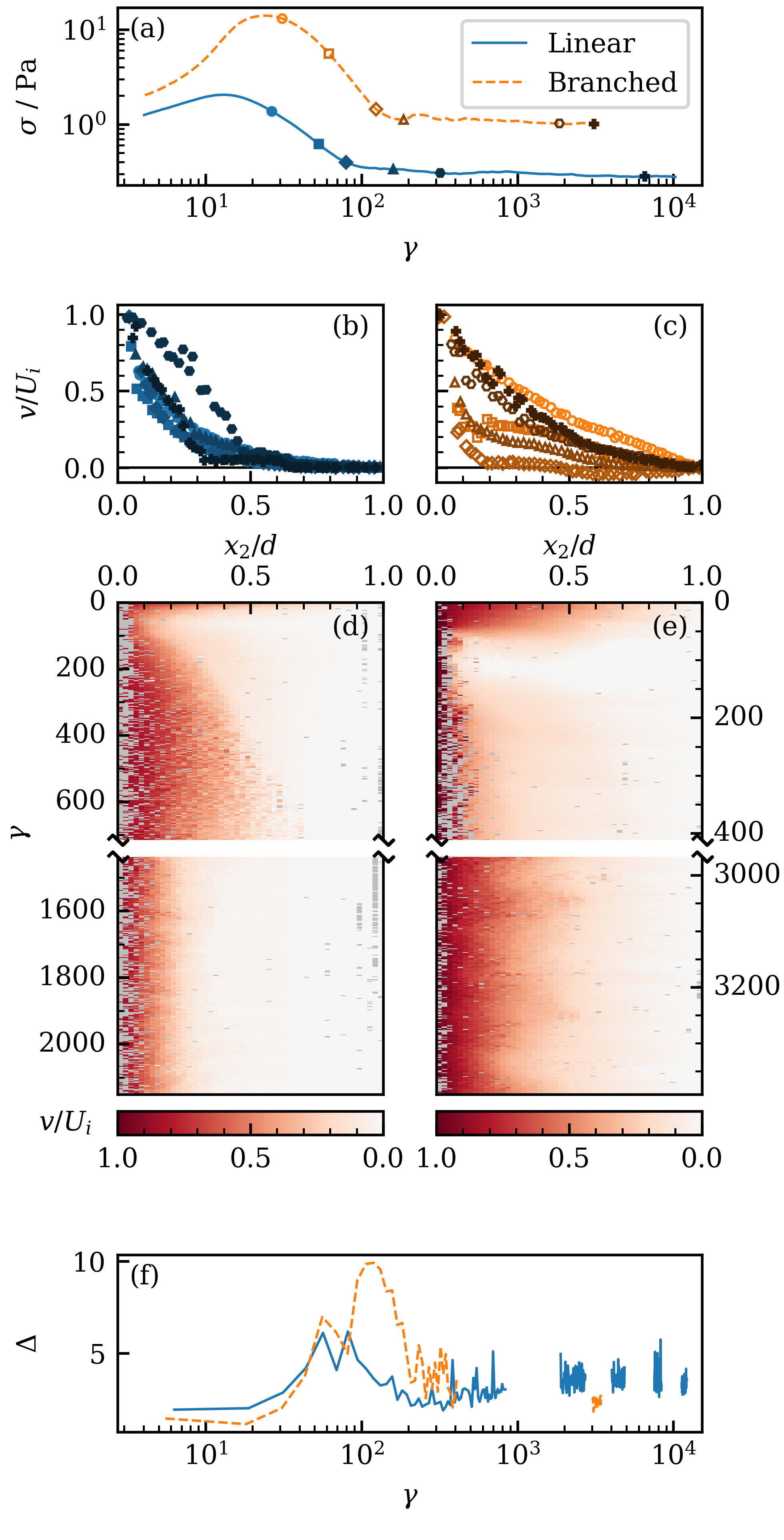}
    \caption{Local velocimetry of linear and branched WLMs at applied Wi = 200: (a)~shear stress evolution with shear strain; (b,c)~selected velocity profiles for (b)~linear and (c)~branched WLMs \add{with shear strain indicated by the matching symbol in (a)}; (d,e)~maps of local velocity evolution with shear strain \add{for linear (d) and branched (e) micellar solutions}; (f)~evolution of the shear rate inhomogeneity parameter $\Delta$ with shear strain.}
    \label{fig:ptv200}
\end{figure}
Fig.~\ref{fig:ptv200} shows the evolution of the flow for linear and branched WLMs at Wi = 200. Although the general trends of the flow evolution are the same between Wi = 50 and Wi = 200, there are some differences. For example, the maximum in flow inhomogeneity, $\Delta_{\mathrm{max}}$, increases, and the strain at which the maximum in inhomogeneity occurs shifts to higher values. In addition, the flow of the branched WLMs develops a wavy velocity profile during the stress decay period. These fluctuations could be linked to instabilities, which will be assessed in the following section. The waves in the velocity profile of the branched system soon fade with $\Delta$ reaching a quasi-steady value around $\gamma \approx 300$, and the quasi-steady velocity profile features a broad \add{(shear-thinning)} curve similar to that for Wi~=~50.

The evolution of the flow profiles at smaller Wi within the stress plateau follow similar trends to those shown at Wi = 50 and Wi = 200 (see Fig.~S9 in the supplementary materials that shows the evolution of the flow at Wi~=~5). At higher Wi that are beyond the end of stress plateau, both linear and branched wormlike micellar solutions show chaotic velocity profile evolution suggesting highly unstable flow conditions (see Fig.~S10 in the supplementary materials).

Fig.~(\ref{fig:DeltaWi}) shows the degree of flow inhomogeneity in linear and branched micellar solutions as a function of imposed Wi using $\Delta_\mathrm{max}$ and $\Delta_\mathrm{s/s}$. Here, $\Delta_\mathrm{max}$ and $\Delta_\mathrm{s/s}$ correspond to the maximum and the steady state shear rate inhomogeneity values. Several observations can be made and are discussed here. First, for the branched micellar solution, $\Delta_\mathrm{s/s}$ does not appreciably increase as Wi increases from below the onset of stress plateau to Wi values that are within the stress plateau region. On the other hand, at this range of Wi, $\Delta_\mathrm{max}$ is always larger than $\Delta_\mathrm{s/s}$ indicating that the degree of inhomogeneity decreases as flow approaches \add{the} quasi-steady limit. These results confirm the above assertion that while the branched micellar solution may \add{develop strong shear thinning, the thinning fades to a more homogeneous structure} during quasi-steady flows for Wi within the stress plateau range. At Wi beyond the end of stress plateau (e.g., Wi = 500), $\Delta_\mathrm{s/s}$ increases dramatically and approaches $\Delta_\mathrm{max}$; this is likely due to formation of strong \add{elastic} instabilities and will be discussed later. For linear wormlike micellar solutions at Wi values below the onset of the stress plateau, $\Delta_\mathrm{s/s}$ is slightly lower than the calculated values for the branched micelles. Within the stress plateau region, the calculated $\Delta_\mathrm{s/s}$ is always larger in the linear wormlike micellar solution than the branched system; additionally, as Wi increases, unlike branched micelles, both $\Delta_\mathrm{max}$ and $\Delta_\mathrm{s/s}$ decay. Note that for imposed Wi close to the onset of stress plateau (e.g., Wi = 2 and Wi = 5), the width of the high shear band is very thin (below the spatial resolution of our apparatus) in linear wormlike micelles; therefore, $\Delta_\mathrm{max}$ and $\Delta_\mathrm{s/s}$ cannot be accurately determined.
\begin{figure}[ht]
    \centering
    \includegraphics[width=7cm]{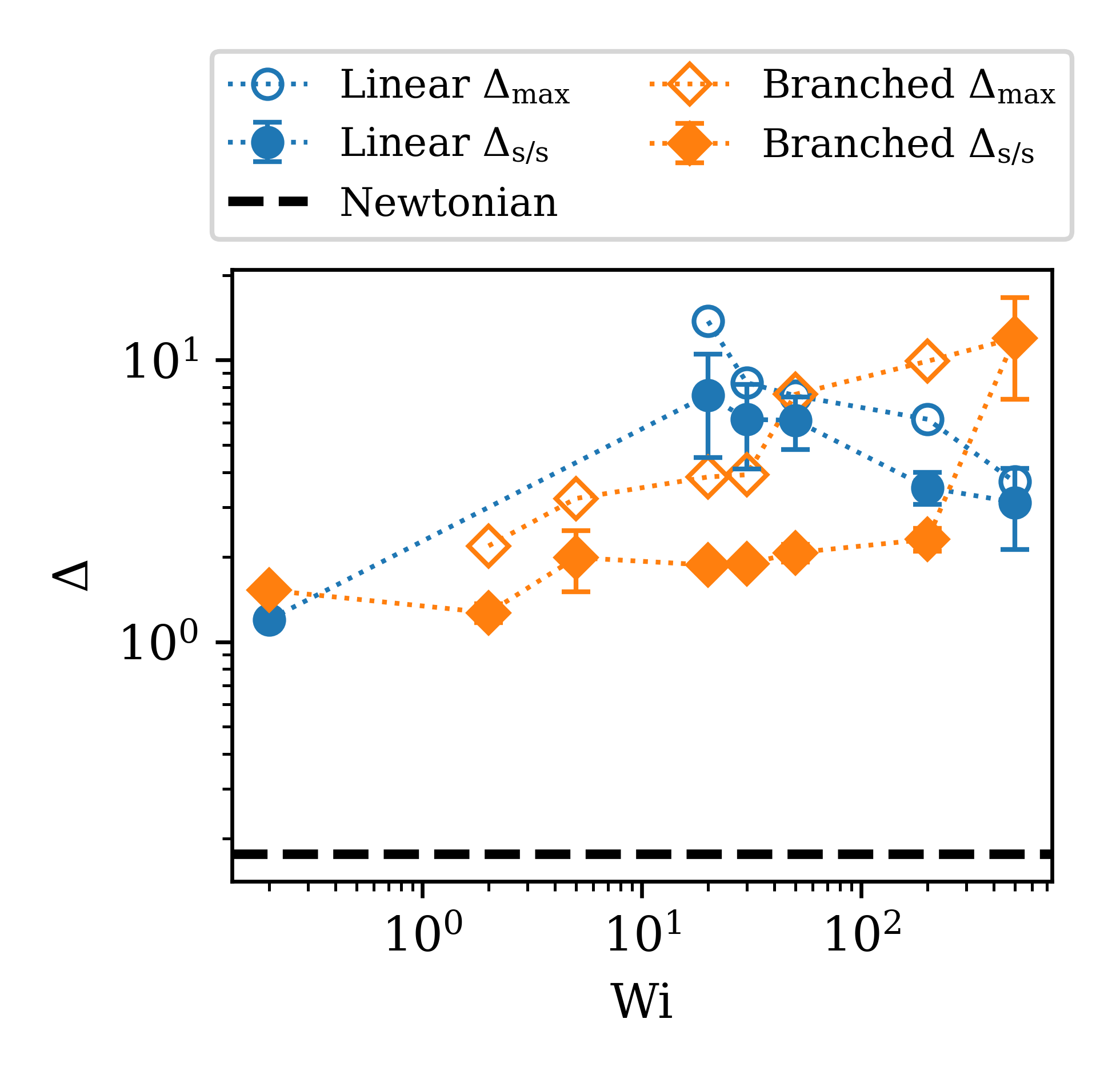}
    \caption{Summary of shear rate inhomogeneity $\Delta$ through the shear stress plateau. Empty symbols represent the maximum value of $\Delta$ observed during the overshoot, and filled symbols represent the quasi-steady average with error bars showing fluctuation. Dotted lines connecting markers are a visual aid.}
    \label{fig:DeltaWi} 
\end{figure}

\subsection{Flow-induced birefringence}

\begin{figure}[ht]
    \centering
    \includegraphics{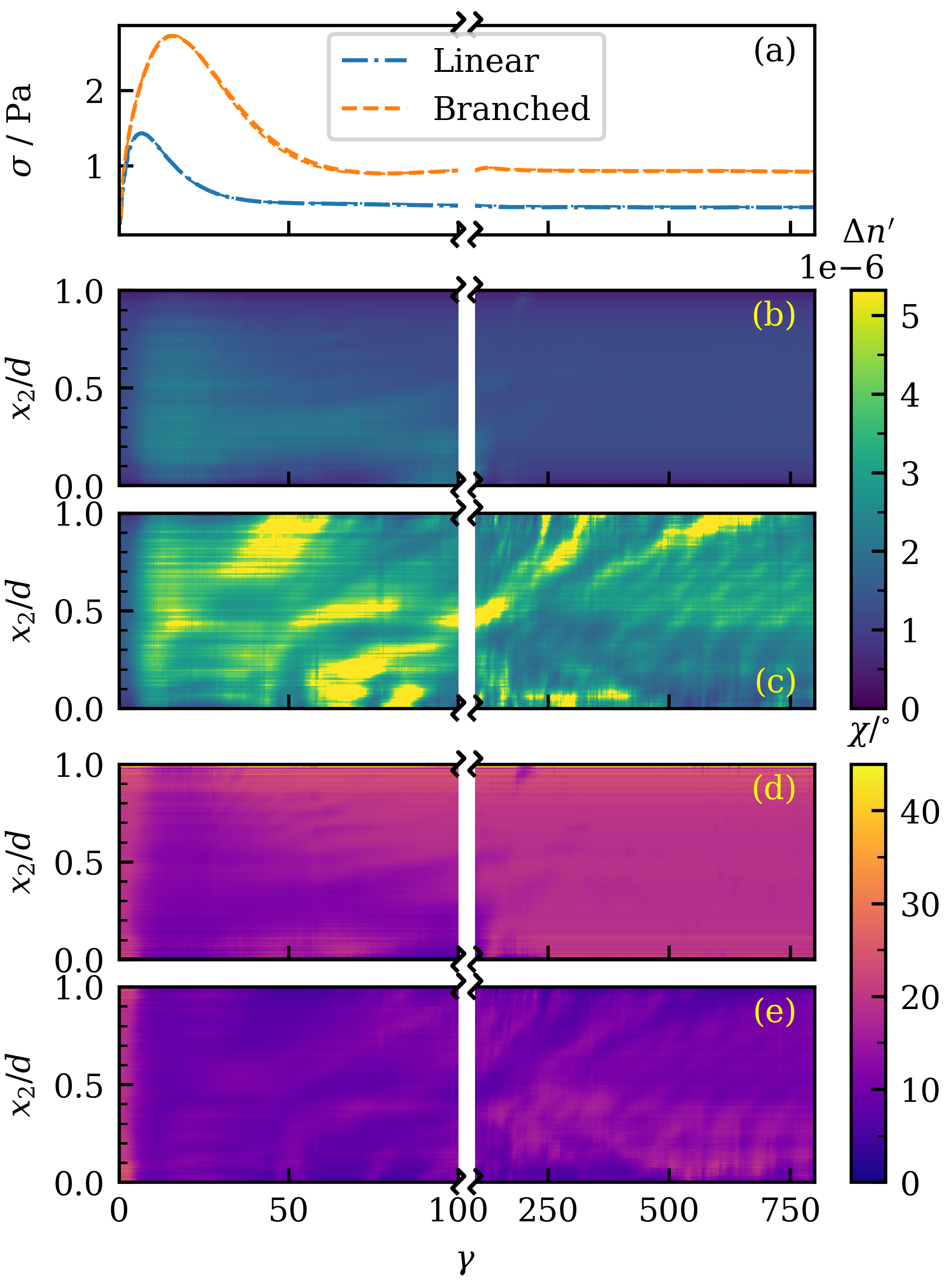}
    \caption{Flow-induced birefringence of linear and branched WLMs at applied Wi = 50: (a)~shear stress evolution with shear strain of (L)~linear and (B)~branched WLMs at both polarizer orientations needed for the Osaki method, (b,c)~spatially and temporally resolved birefringence for (b)~linear and (c)~branched WLMs, (d,e)~spatially and temporally resolved extinction angle for (d)~linear and (e)~branched WLMs.}
    \label{fig:fib50}
\end{figure}

For Wi $\le$ 30, flow-induced birefringence is weak and cannot be detected in our device. However, for Wi $>$ 30, flow-induced birefringence becomes more pronounced. For example, Fig.~\ref{fig:fib50} shows the birefringence profiles for linear and branched WLMs at Wi~=~50 in terms of both birefringence magnitude ($\Delta n^{\prime}$) and extinction angle ($\chi$). For linear WLMs following imposition of the flow, birefringence appears across the gap of the TC cell during the shear stress overshoot. After the stress overshoot and during the stress decay period, a birefringent band is observed near the inner cylinder of the TC cell that occupies approximately 40~\% of the gap width, which is close to the width of the high shear band (30~\% of the gap) under quasi-steady flow observed in the rheo-PTV experiments (Fig.~\ref{fig:fib50}(b)). 
Associated with the birefringent band is a band with low values of the extinction angle $\chi$ indicating micellar alignment in the high shear band near the inner cylinder of the TC cell (Fig.~\ref{fig:fib50}(d)). For $\gamma \geq 100$, the apparent birefringence weakens in the linear wormlike micellar solution. On the other hand, for branched WLMs at this Wi, there is a much stronger birefringence and a stronger alignment during the shear stress overshoot (see Fig.~\ref{fig:fib50}(c) and (e)). This observation is similar to those of Wu $\&$ Mohammadigoushki that showed the birefringence intensity is stronger in branched micellar solutions than the linear wormlike micelles in flow past a falling sphere~\cite{Wu2021}. As time progresses, around $\gamma \approx 25$, the observed birefringence and the extinction angle in the branched micellar system become chaotic. This chaos is likely due to strong \add{elastic} instabilities in the flow field that cause the light to be bent along the $x_3$ direction; we provide a detailed discussion of instabilities in subsection~\ref{ssn:thetaz}. 

\begin{figure}[ht]
    \centering
    \includegraphics{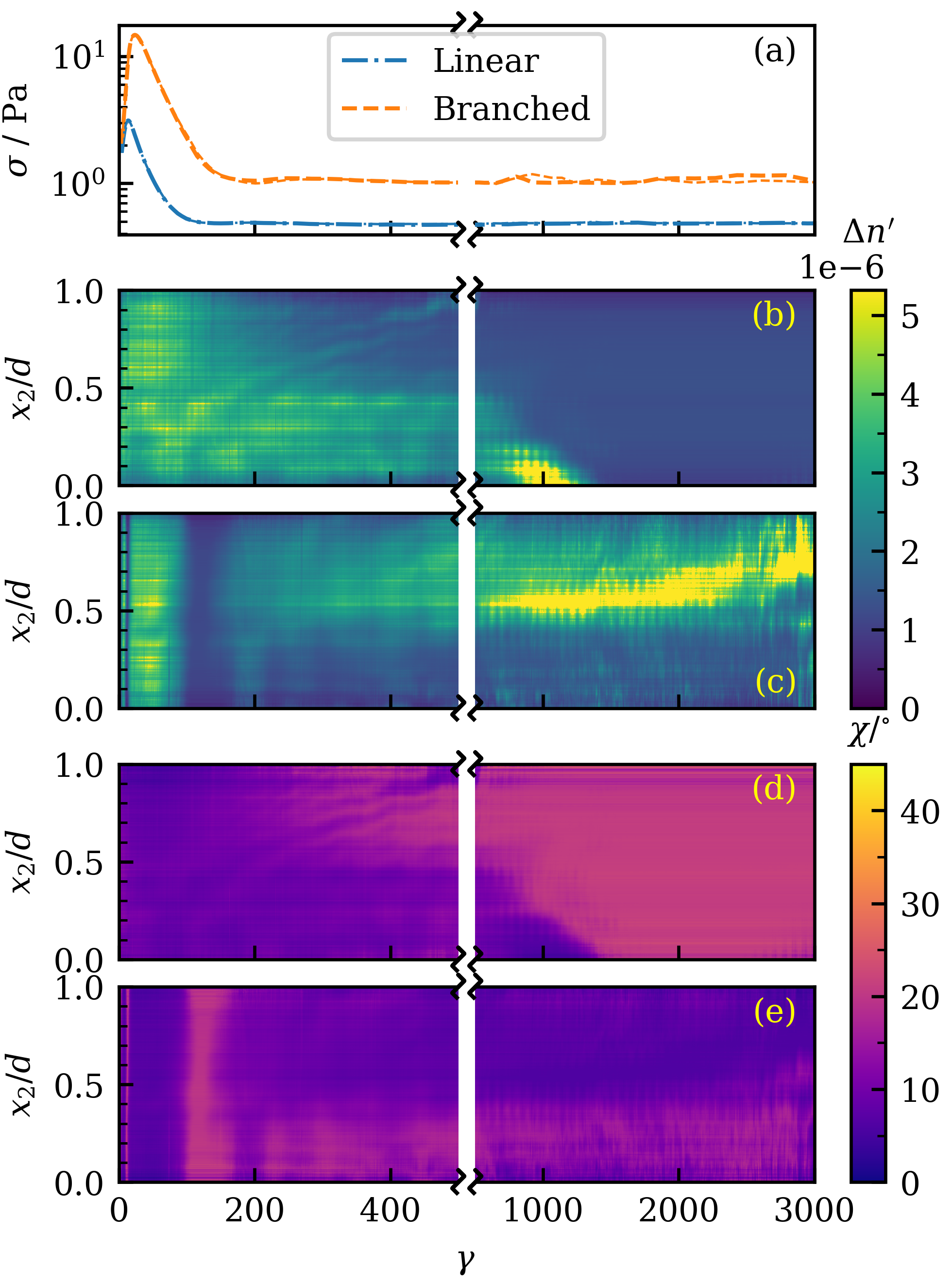}
    \caption{Flow-induced birefringence of linear and branched WLMs at applied Wi = 200: (a)~shear stress evolution with shear strain of linear and branched WLMs at both polarizer orientations needed for the Osaki method, (b,c)~spatially and temporally resolved birefringence for (b)~linear and (c)~branched WLMs, (d,e)~spatially and temporally resolved extinction angle for (d)~linear and (e)~branched WLMs.}
    \label{fig:fib200}
\end{figure}

At Wi~=~200, the bireringence for linear WLMs  shown in Fig.~\ref{fig:fib200} (b.d) is qualitatively similar to those shown for Wi = 50. There is uniform birefringence from the start of applied shear to the end of the shear stress overshoot period ($\gamma \le 100$), beyond which a low-birefringence region forms near the outer cylinder. The width of this low birefringent band is approximately the same as the size of the low shear band data of Fig.~\ref{fig:ptv200}(b). Around $\gamma \approx 1000$, the light in the high birefringence region becomes distorted and ultimately disappears. It is possible that at such high strains, secondary flows and instabilities may form in the high shear band that may in turn distort light and interfere with rheo-FIB measurements. We will come back to this observation in subsection~\ref{ssn:thetaz}. 

\subsection{Visualization of turbidity in the $x_2$-$x_3$ plane}

\begin{figure}[h]
    \centering
    \includegraphics{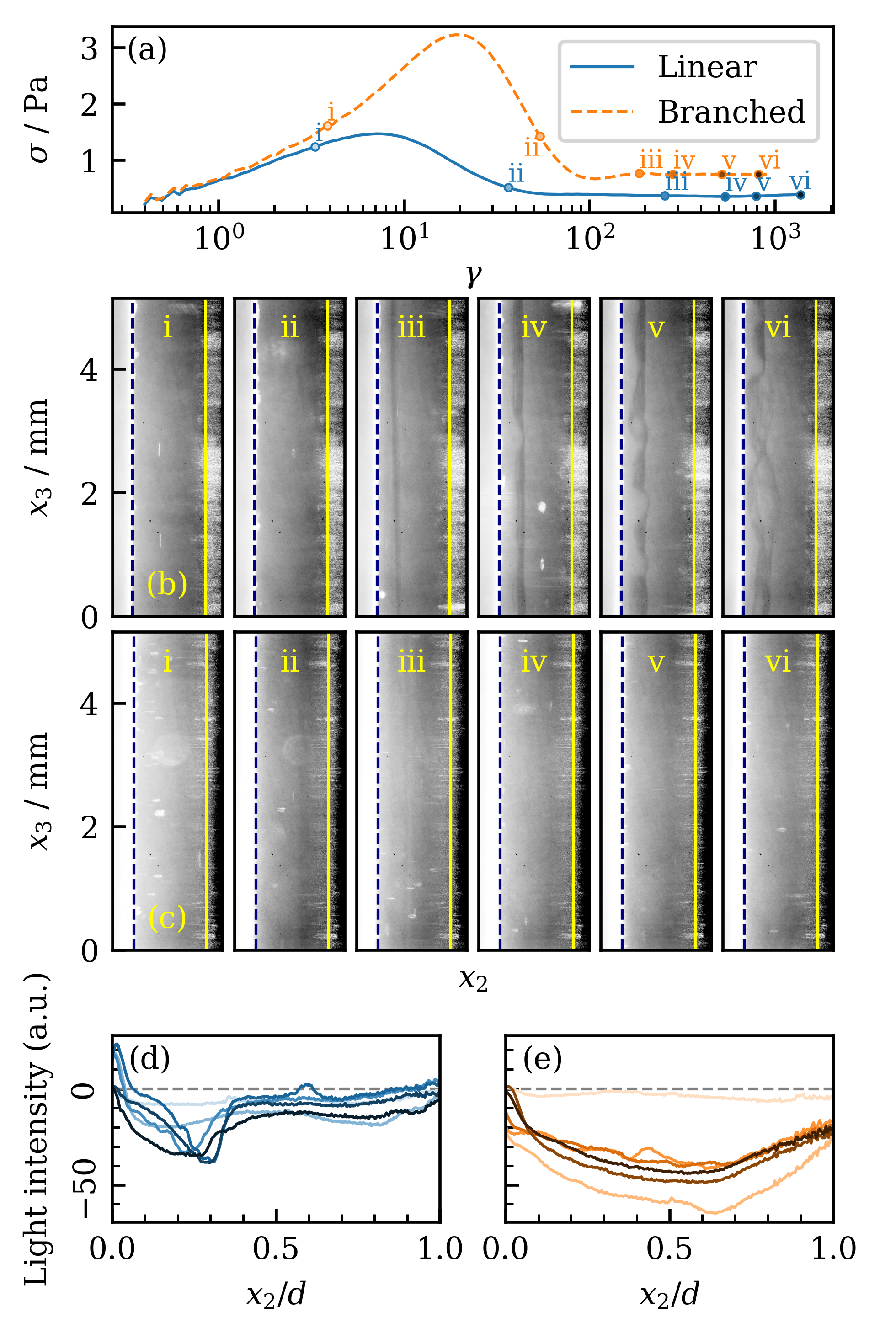}
    \caption{Turbidity visualized by illuminating the $x_2-x_3$ plane for linear and branched WLM solutions at Wi~=~50. (a)~Shear stress as a function of shear strain. \add{(b,c)~}Snapshots showing the evolution of turbidity for linear (b) and branched (c) WLMs at selected shear strains. \add{The} outer cylinder is marked with a solid yellow line, and the inner cylinder is marked with a dashed blue line. Averaged turbidity intensity as a function of radial position within the gap of the TC cell \add{is} shown for \add{the} linear (d) and \add{the} branched (e) micellar system. The dashed horizontal line indicates the initial state, and as shear strain progresses\add{,} the curve color changes from lighter to a darker color.}
    \label{fig:rz50}
\end{figure}

Under shear banding flows, some WLM solutions develop turbidity in the high shear band. Lerouge and co-workers~\cite{Fardin2012a,Lerouge2008} and Mohammadigoushki and Muller~\cite{Mohammadigoushki2016} imaged turbidity in the vorticity ($x_2$-$x_3$) plane in a shear banding solution of CTAB/NaNO$_3$ under a startup of a simple shear flow. These authors observed the development of the shear banding and the onset of elastic instabilities, with the high shear band clearly visible as a bright (turbid) region near the inner cylinder of the TC cell. In the same manner, we have examined the evolution of turbidity in our flows of linear and branched micellar solutions to assess if there are any differences between the turbidity of these samples. Fig.~\ref{fig:rz50} shows the evolution of shear stress and the turbidity profiles in the $x_2-x_3$ plane for linear and branched WLMs at Wi~=~50.\par
For linear WLMs, the fluid is initially turbid with a turbidity that is uniform throughout the gap of the TC cell (see the dashed horizontal line in Fig.~\ref{fig:rz50}(d)). As time progresses, a region of low turbidity emerges near the inner cylinder of the TC cell (see also movie 1 in the supplementary materials). As the shear stress reaches the steady value, the width of the lower turbidity region is around $x/d$ = 0.2, in agreement with formation of a high shear band near the inner cylinder in Fig.~\ref{fig:ptv50}(b,d). As time continues to progress and for $\gamma \geq 500$, the flow in the high shear band develops instabilities in the linear WLM solution (see v and vi in Fig.~\ref{fig:rz50}(b)). On the contrary, the branched WLM solution appears to be much more stable at Wi=50. The turbidity of the fluid throughout the entire flow field decreases during the stress overshoot. Around $\gamma \approx 200$, some turbidity reemerges near the inner cylinder, but over time, that turbidity fades, and there is no interface between turbid and less turbid regions (see Fig.~\ref{fig:rz50} (c,e)). The latter result is consistent with the curved quasi-steady velocity profile of Fig.~\ref{fig:ptv50}(c) that shows no signs of shear banded velocity profile across the gap of the TC cell for the branched WLM solution.

\begin{figure}[h]
    \centering
    \includegraphics{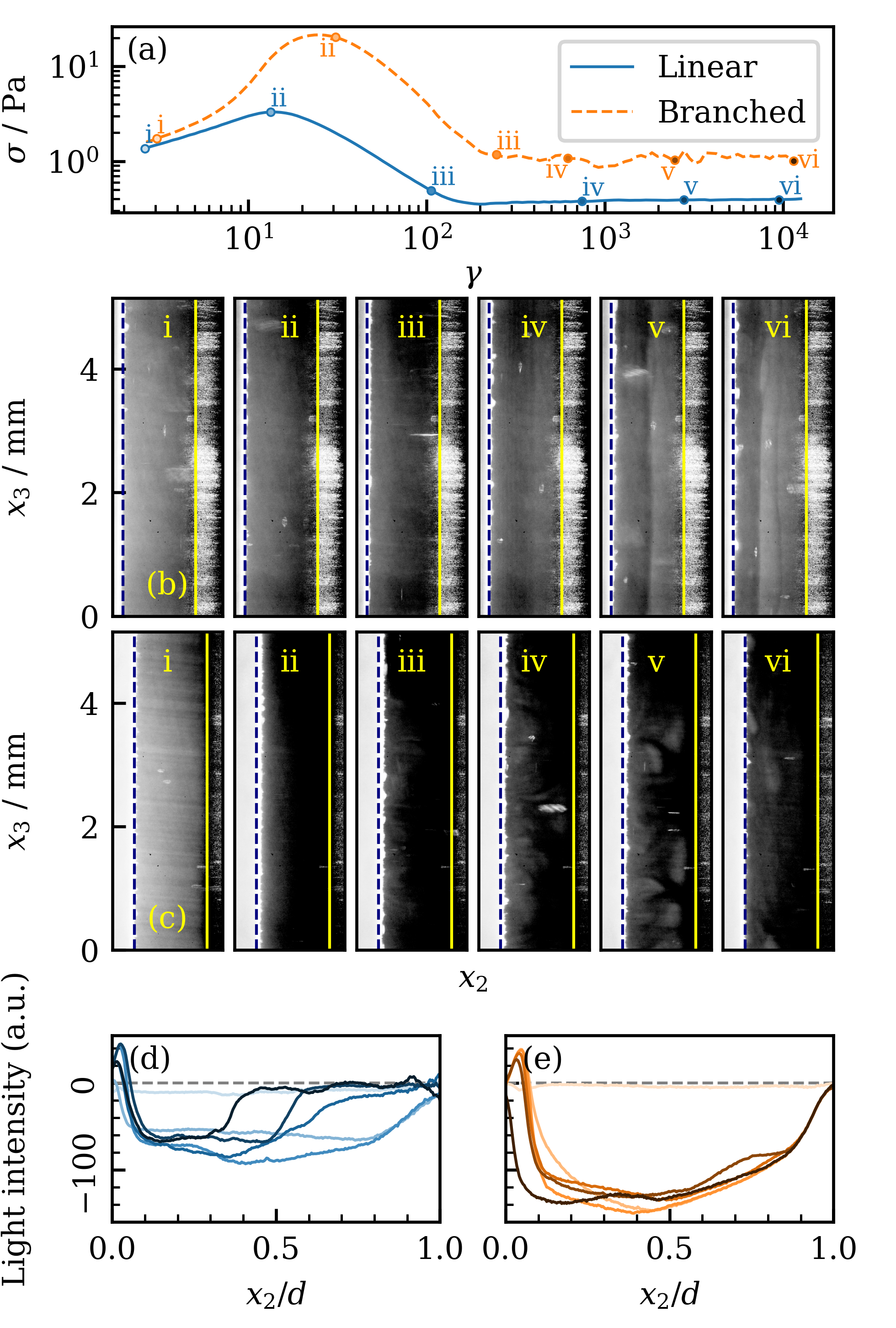}
    \caption{Turbidity visualized by illuminating the $x_2-x_3$ plane for linear and branched WLM solutions at Wi~=~200. (a)~Shear stress as a function of shear strain. \add{(b,c)~}Snapshots showing the evolution of turbidity for linear (b) and branched (c) WLMs at selected shear strains. \add{The} outer cylinder is marked with a solid yellow line, and the inner cylinder is marked with a dashed blue line. Averaged turbidity intensity as a function of radial position within the gap of the TC cell \add{is} shown for \add{the} linear (d) and \add{the} branched (e) micellar system. The dashed horizontal line indicates the initial state, and as shear strain progresses\add{,} the curve color changes from lighter to a darker color.}
    \label{fig:rz200}
\end{figure}
At a larger imposed Weissenberg number, Wi = 200, the evolution of the turbidity in linear WLMs is generally similar to that at Wi~=~50 with a few differences. The initial drop in turbidity during the shear stress overshoot is stronger for both linear and branched micellar solutions (the same light intensity, exposure time, brightness, and gamma adjustment were used for all images in Figs.~\ref{fig:rz50} and \ref{fig:rz200}). For the linear sample, the turbidity interface initially forms at $x/d \approx 0.5$ (see Fig.~\ref{fig:rz200}(b)(v) and Fig.~\ref{fig:rz200}(d)), which is consistent with the rheo-PTV results shown in Fig.~\ref{fig:ptv50}(d). As the flow evolves to the quasi-steady state, the turbidity interface migrates to $x/d\approx $  0.35 (see Fig.~\ref{fig:rz200}(b,vi) and (d)). Additionally, a careful inspection of the quasi-steady turbidity image of Fig.~\ref{fig:rz200}(b,vi) reveals formation of a highly unstable thin region that is sandwiched between the high and low turbidity bands (see also movie 2 in the supplementary materials). In contrast, the branched WLM solution at applied Wi~=~200 shows markedly different evolution from lower Wi. Upon imposition of the shear, the sample is turbid throughout the gap of the TC cell. At larger strains, there is a strong drop in turbidity across the gap, and the flow field remains dark until $\gamma \approx 800$ (see Fig.~\ref{fig:rz200}(c) (i-iv)). At that point, instabilities appear in the flow field. These instabilities are strongest near the inner cylinder but can sometimes occupy most or all of the gap, as seen in Fig.~\ref{fig:rz200}(c) (v). This is consistent with Fig.~\ref{fig:ptv200}, where the quasi-steady flow features some instabilities near the inner cylinder but no distinct shear banding interface. 

\begin{figure}[h]
    \centering
   \includegraphics[width=9cm]{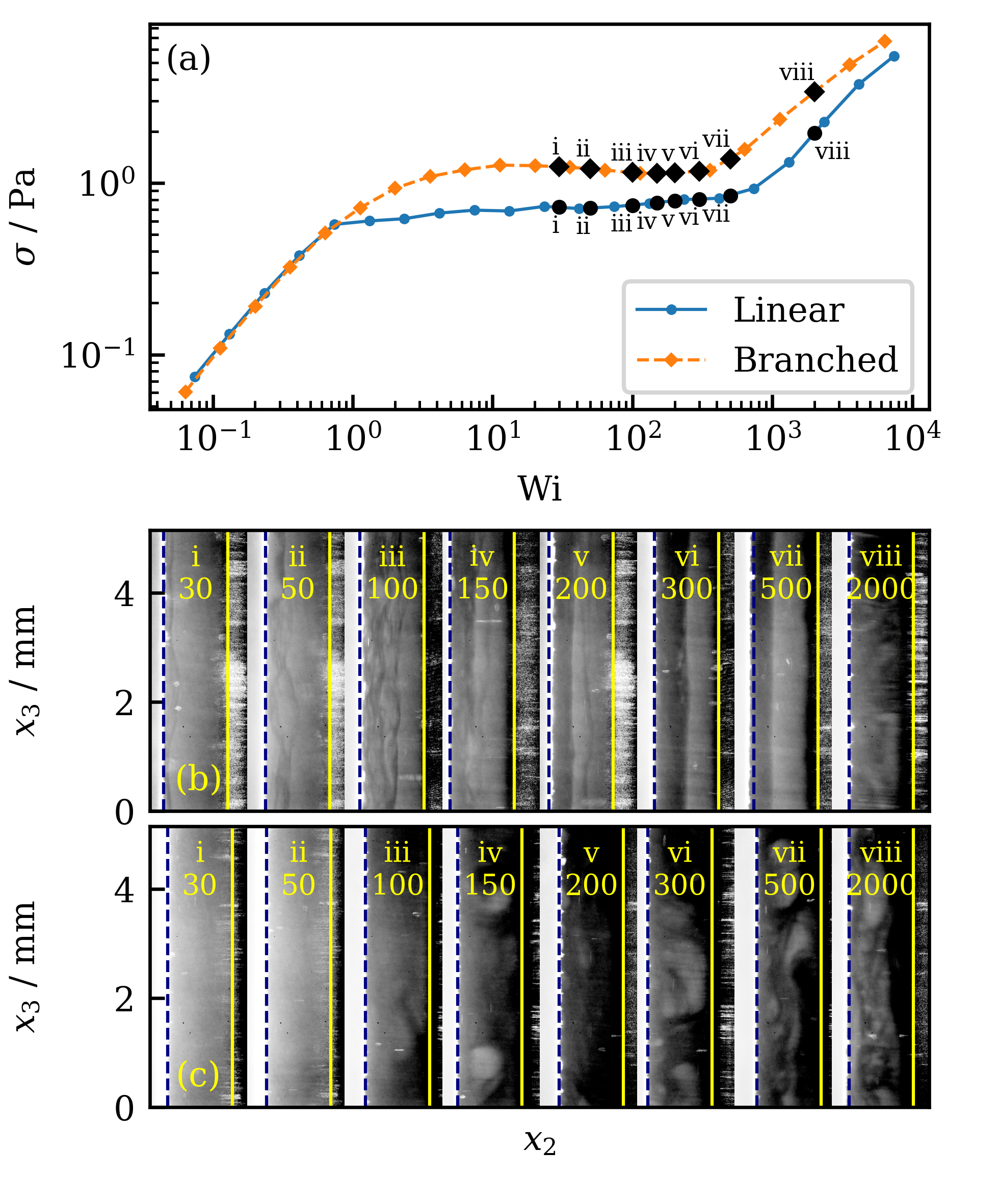}
    \caption{Turbidity visualized by illuminating the $x_2-x_3$ plane for linear and branched WLMs at quasi-steady-state. (a)~Flow curves showing selected Wi. (b)~Turbidity visualized for linear WLMs at selected Wi. (c)~Turbidity visualized for branched WLMs at selected Wi.}
    \label{fig:rzss}
\end{figure}

The quasi-steady turbidity profiles of linear and branched systems are shown in Fig.~\ref{fig:rzss} for different imposed Wi. For the linear WLM solution at low Wi (e.g., Wi $< 30$), the width of the high shear band (less turbid region) is thinner than the spatial resolution of our optical device. Therefore, such profiles are not included here. At Wi~=~30, a thin and unstable band with a slightly lower turbidity than the rest of the gap, forms near the inner cylinder (see Fig.~\ref{fig:rzss}(b,i)). As Wi increases, the width of this highly unstable band increases. For Wi $>100$, a dark phase replaces the unstable band near the inner cylinder of the TC cell and the unstable band is now sandwiched between the two dark and bright bands (see Fig.~\ref{fig:rzss}(b,iv,v). At still higher Wi (e.g., Wi = 300 or 500), the unstable interface between the two bands disappears until the flow becomes chaotic at Wi = 2000, which is beyond the end of the stress plateau. While the linear WLM system shows signatures of a typical shear banded system, the quasi-steady images of the branched solution show a remarkably different behavior. At low Wi (Wi $\leq$ 100), the turbidity contrast is fairly uniform within the gap of the TC cell with no signs of turbidity contrast that is expected for shear banding systems. For Wi $>100$, the flow becomes unstable and remains unstable at higher Wi values. Clearly, the instabilities in the branched system have a different nature than those observed in the high shear band of the linear wormlike micellar system. 

\subsection{\add{Elastic} Instabilities}
\label{ssn:thetaz}
\begin{figure*}[ht]
    \centering
    \includegraphics{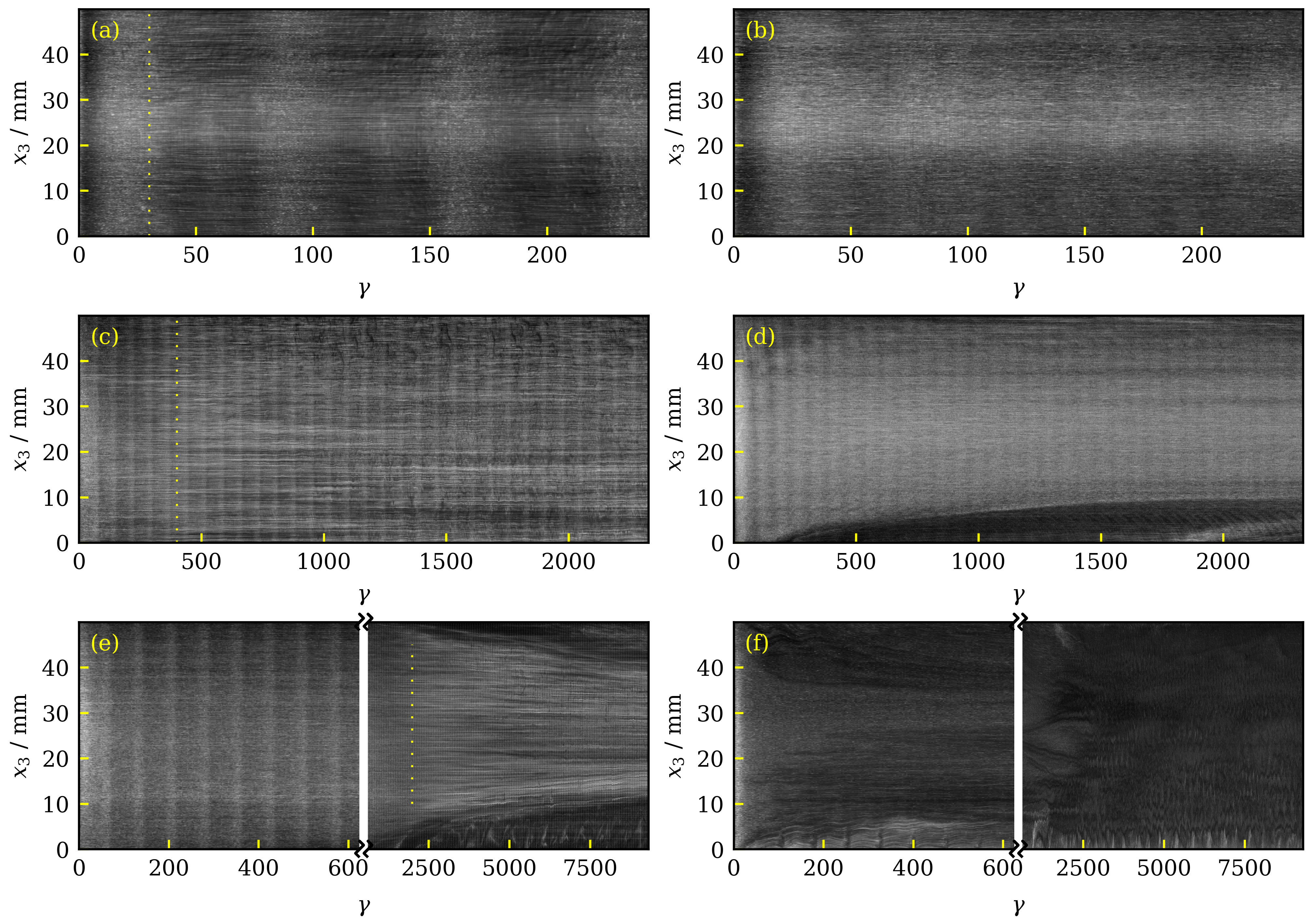}
    \caption{Spatio-temporal evolution of flows visualized with mica flakes for linear (left column; a,c,e) and branched (right column; b,d,f) WLM solutions. From top to bottom, plots correspond to Wi~=~5 (a,b), Wi~=~50 (c,d) and Wi~=~200 (e,f). The vertical dashed lines mark the onset of secondary flow formation.}
    \label{fig:thetaz5}
\end{figure*}

Fig.~(\ref{fig:thetaz5}) shows the spatio-temporal evolution of the secondary flows of the linear (Fig.~\ref{fig:thetaz5}(a,c,e)) and branched (Fig.~\ref{fig:thetaz5}(b,d,f)) micellar solutions following a startup of steady shear flow to three different Wi values in the shear stress plateau. Although the flow of the branched micellar solution remains stable at Wi~=~5 (Fig.~\ref{fig:thetaz5}(b)), weak secondary flows appear at $\gamma \approx 30$ as thin and spatially invariant stripes for the linear wormlike micellar system, and are shown in Fig.~\ref{fig:thetaz5}(a). Note that the small striations in Fig.~\ref{fig:thetaz5}(b) are due to the finite size of the mica flakes and are present even at Wi~=~0.2 (see also Fig.~S11 in the supplementary materials). In addition, the vertical black and white stripes in these figures are caused by spatial inhomogeneities of the rotating inner cylinder of the TC cell. At higher imposed Wi values, the high shear band migrates towards the outer cylinder and secondary flows become more visible. In Fig.~\ref{fig:thetaz5}(c,d) the spatio-temporal evolution of the secondary flows of the linear and branched micellar solutions are shown at Wi~=~50. For the linear WLM solution, secondary flows begin to appear at $\gamma \approx 400$, which is in agreement with the onset of instabilities observed in Fig.~\ref{fig:rz50}(b). Unlike the linear WLM solution, the flow of the branched system remains stable through $\gamma > 2000$ except near the ends (see Fig.~\ref{fig:thetaz5}(d)). At the lower end of the TC cell, there is a strong deviation from stable flow starting around $\gamma \approx 100$. This unstable region grows to 10 mm above the bottom of the TC cell until $\gamma \approx 1500$, beyond which the unstable region has stopped growing. Additionally, some instabilities are observed near the upper end of the TC cell. As noted before, the rheo-PTV and turbidity results were obtained on fluid near the center far away from either end of the TC cell. Therefore, the non-shear banded velocity profiles and the absence of a visible turbidity contrast in branched wormlike micelles at low and intermediate Wi values are not caused by these end effects. On the other hand, the chaotic FIB pattern observed for branched micellar system at Wi = 50 in Fig.~\ref{fig:fib50}(c,e)) is presumably caused by the end effects because the light beam must pass through the entire height of the TC cell. Unstable flows of branched solutions anywhere in the gap (particularly at both ends of the TC cell) may bend the light and cause it to fall on the screen at a different location from where it otherwise would. Therefore, chaotic end effects that bend the light form a chaotic pattern for $\Delta n^{\prime}$ and $\chi$.

At larger imposed Weissenberg numbers (e.g. Wi~=~200; see Fig.~\ref{fig:thetaz5}(e,f)), strong end effects become visible for both linear and branched WLM solutions at different extents and time scales. For the linear WLM system, spatially invariant secondary vortices and end effects emerge simultaneously around $\gamma \approx 1500$. These instabilities become clearly visible around $\gamma \approx 2000$. The emergence of such strong instabilities may explain the breakdown of visible birefringence banding in Fig.~\ref{fig:fib200}(b) around $\gamma \approx 1000$. The instabilities spanning the center of the TC cell then intensify somewhat, and the instabilities caused by end effects grow significantly through $\gamma \approx 8000$ and beyond. For the branched WLM solution, end regions with chaotic flow emerge almost immediately after the initial application of the steady shear rate. They then grow towards the center, and by $\gamma \approx 2000$, they span the entire TC cell height. At still higher Wi, a similar chaotic pattern is observed for both linear and the branched micellar solutions (see e.g., Fig. S12-S13 of the supplementary materials). \par 

We can deduce from the above results that the branched WLM solution is more susceptible to end effects than the linear WLM solution. Previous rheo-optical measurements on WLM solutions have used a range of TC cell aspect ratios $\Gamma = h/d$ (e.g., 3.33-40\cite{Fardin12b}, 35.4~\cite{Lerouge2008,Fardin2012a}, 60.7~\cite{Mohammadigoushki2016,Mohammadigoushki2017} or 42.3~\cite{Rassolov2020,Mohammadigoushki2019}). For the smallest aspect ratio of $\Gamma = 3.3$,  Fardin and co-workers showed that the temporal evolution of secondary vortices becomes spatially variant along the $x_3$ axis of their TC cell, and therefore, end effects become increasingly important at such small aspect ratios. Conversely, for aspect ratios of $\Gamma \geq 35.4$, end effects have shown to be negligible for a wide range of WLM solutions by various research groups~\cite{Fardin12b,Lerouge2008,Mohammadigoushki2016,Rassolov2020}. Although the aspect ratio of the TC cell used in this study ($\Gamma = 42.3$) was expected to be large enough based on our previous experiments and published literature, the temporal evolution of the secondary vortices shown in Fig.~\ref{fig:thetaz5}(d,e,f) clearly demonstrate that end effects are significant for branched micellar solutions at intermediate and large Wi values. As noted before, these two micellar solutions have similar rheological properties, hence, the sensitivity of the branched micellar solution to the end effects may be related to their microstructure.

\section{Summary and Conclusions}
\label{conclusions}

In summary, we studied the effects of micellar microstructure (linear or branched) on the spatio-temporal evolution of flow of WLM solutions with matched rheology. We found that while both the linear and the branched WLM solutions exhibit a shear stress plateau in the flow curve and undergo strong transient shear banding following the shear stress overshoot, the quasi-steady flow of branched WLMs does not feature shear banding, unlike the linear WLMs. Instead, the quasi-steady flow of branched WLMs is represented by a gently curved velocity profile. We also found that, consistently with prior literature\cite{Wu2021}, branched WLMs produce a stronger birefringence under shear flow. However, the measured birefringence of branched WLMs also features chaos due to end effects. Our turbidity measurements indicated a clear turbidity gradient between the high and low shear bands of the linear WLM solution, with the high band exhibiting a lower turbidity. In contrast, the branched WLM solution exhibits a decrease in turbidity with no turbidity contrast within the gap of the TC cell, thereby, corroborating the results of velocimetry measurements. \add{The above results are observed for a broad range of Wi (e.g., Wi = 5, 10, and 30), where complex flow phenomena are not present. }

\add{At higher Wi values,} elastic instabilities form for both linear and branched WLM solutions towards the end of \add{the} stress plateau ($30 \lesssim \mathrm{Wi} \lesssim 500$ for linear and $100 \lesssim \mathrm{Wi} \lesssim 500$ for branched systems, respectively), but by two different mechanisms. In the linear WLM solution, shear banding develops with initially stable flow, and then instabilities form at the high shear rate band. In branched WLM solutions, there is initially stable flow with no shear banding, and then strong end effects form that at higher Wi (Wi $\gtrsim$ 100) spread axially to span the entire height of the TC cell. These results highlight the significant role of micellar branching on the transient evolution of flow in WLM solutions. \add{We note that the slight differences in rheology and ionic strength of these two WLMs, although unlikely to have a significant effect, may also contribute to the reported differences in flow structures of these two micellar solutions.} 

Finally, our findings suggest that the results from future studies involving TC flows of branched micellar solutions, especially those involving a beam propagating along the vorticity direction (e.g. rheo-FIB, rheo-SANS), should be interpreted with caution due to potential role of end effects. Alternatively, future experiments can be conducted using TC cell geometries with large aspect ratios ($\Gamma > 50$) to reduce the impacts of end effects. Increasing the aspect ratio of our rheo-optical apparatus requires major modifications to it and is the focus of our future studies.

\section{Acknowledgments}
We are grateful to National Science Foundation for financial support of this work through award CBET CAREER 1942150. \add{ NMR experiments were performed at the US National High Magnetic Field Laboratory (NHMFL), which is supported by the State of Florida and the National Science Foundation Cooperative Agreement No. DMR-1644779. We are grateful to Samuel Holder and Samuel Grant for assisting us with the magnet time during their scheduled experiments. }

\bibliography{refs}
\end{document}